\documentclass[aip,jcp,twocolumn,nobibnotes,reprint,groupedaddress]{revtex4-1} 

\usepackage{graphicx,amsfonts,amsmath,amsbsy,amssymb,hyperref,subfigure,bbold,placeins,xspace}
\usepackage[american]{babel}
\newcommand{\refeq}[1]{{Eq.~(\ref{#1})}}
\newcommand{\reffig}[1]{{Fig.~\ref{#1}}}
\newcommand{\refsec}[1]{{Sec.~\ref{#1}}}

\DeclareMathOperator{\Tr}{Tr}

\hypersetup{breaklinks,colorlinks=true,linkcolor=blue,citecolor=blue,urlcolor=blue,filecolor=blue}

\newcommand{\contributions}{{
WZV, LW, and JJS contributed to the development of the machine learning/GPR ideas presented here.
WZV wrote the code used for all of the GPR calculations presented here with advice from LW.
GS, SM, and TD separately and together provided investigations of cubic splines which were important in establishing benchmark comparisons for GPR. These were subsequently reproduced by code written by WZV. 
WZV and GS provided mentoring of other students. 
WZV collected all of the DMQMC data, as part of this current investigation, and during a separate project written up elsewhere.\cite{vanbenschoten_piecewise_2022}
WZV and JJS wrote the manuscript with contributions from TD (cubic splines methods) and LW (GPR methods).
All authors provided analysis of data, including its interpretation. 
All authors checked and confirmed the accuracy of the final manuscript, including the author list. 
All work was performed at the University of Iowa.
}}

\begin{document}

\title{Electronic specific heat capacities and entropies from density matrix quantum Monte Carlo using Gaussian process regression to find gradients of noisy data}

\author{William~Z.~Van Benschoten}
\author{Laura Weiler}
\author{Gabriel~J.~Smith}
\author{Songhang Man}
\author{Taylor DeMello}
\author{James~J.~Shepherd}
\email{james-shepherd@uiowa.edu}
\affiliation{Department of Chemistry, University of Iowa}
\date{\today}

\begin{abstract}

We present a machine learning approach to calculating electronic specific heat capacities for a variety of benchmark molecular systems.
Our models are based on data from density matrix quantum Monte Carlo, which is a stochastic method that can calculate the electronic energy at finite temperature. 
As these energies typically have noise, numerical derivatives of the energy can be challenging to find reliably.
In order to circumvent this problem, we use Gaussian process regression to model the energy and use analytical derivatives to produce the specific heat capacity. 
From there, we also calculate the entropy by numerical integration. 
We compare our results to cubic splines and finite differences in a variety of molecules whose Hamiltonians can be diagonalized exactly with full configuration interaction. 
We finally apply this method to look at larger molecules where exact diagonalization is not possible and make comparisons with more approximate ways to calculate the specific heat capacity and entropy.

\end{abstract}

\maketitle

\section{Introduction}

The electronic specific heat capacity and electronic entropy are known to play a significant role in the phase transitions of solid state materials, such as metal oxides\cite{zhou_configurational_2006,paras_electronic_2020}, pure metals\cite{harrison_phase_2019}, and complex mixtures of metals and metalloids\cite{wang_complex_2016}. 
Furthermore, material properties such as conductivity and superconductivity are related to the electronic specific heat capacity\cite{paglione_high-temperature_2010} and electronic entropy\cite{fong_measurement_2013} of the materials.
Because of the impact that the electronic specific heat capacity and electronic entropy can have on a system's behavior, the need for electronic structure calculations that accurately model these quantities is important.

There are a lot of finite temperature electronic structure methods which have been continuously developed in the past couple of decades, and these methods span a broad range of accuracy and computational cost.
Some examples of low cost methods applicable to large systems are finite temperature density functional theory\cite{ellis_accelerating_2021,cytter_stochastic_2018,pittalis_exact_2011,karasiev_generalized-gradient-approximation_2012,eschrig_t_2010,pribram-jones_thermal_2016,smith_exact_2016} and finite temperature Hartree--Fock (HF)\cite{he_finite-temperature_2014,hermes_finite-temperature_2015,sjostrom_temperature-dependent_2012}.
Several theories introduce finite temperature perturbation corrections to mean-field methods such as HF to increase the accuracy of their descriptions, but come with an added computational cost. These methods include Green's function\cite{li_sparse_2020,pokhilko_evaluation_2021,harkonen_many-body_2020,yeh_fully_2022,kananenka_efficient_2016,welden_exploring_2016,kas_finite_2017,karrasch_finite-temperature_2010,neuhauser_stochastic_2017,gu_generalized_2020} and many-body perturbation theory\cite{hirata_finite-temperature_2020,he_finite-temperature_2014,santra_finite-temperature_2017,jha_finite-temperature_2020,hirata_finite-temperature_2021,pokhilko_iterative_2022}. 
Next in line are embedding theories, which separate the system into two subsystems: one subsystem that is treated approximately, and another subsystem that is treated using a higher-accuracy method, such as perturbative methods or fully-correlated methods.\cite{sun_finite-temperature_2020,liebsch_finite-temperature_2009,liebsch_temperature_2012,strand_dynamical_2011,knizia_density_2013,kretchmer_real-time_2018,tran_using_2019,cui_efficient_2020,zhai_low_2021,tsuchimochi_density_2015,bulik_electron_2014,zgid_finite_2017,lan_generalized_2017,tran_spin-unrestricted_2018,rusakov_self-energy_2019}
Finite temperature coupled cluster\cite{hummel_finite_2018,white_finite-temperature_2020,harsha_thermal_2022,dzhioev_superoperator_2015,hermes_finite-temperature_2015,shushkov_real-time_2019,white_time-dependent_2019,peng_conservation_2021} methods and selected configuration interaction\cite{harsha_thermofield_2019,harsha_wave_2020} methods allow for more control in the accuracy of the calculation by truncating the methods to approximations based on the number of excitations to include.
The most accurate but most costly methods are the quantum Monte Carlo (QMC)\cite{blunt_density-matrix_2014,malone_interaction_2015,blunt_krylov-projected_2015,yilmaz_restricted_2020,dornheim_ab_2021,brown_path-integral_2013,liu_ab_2018,shen_finite_2020,he_finite-temperature_2019,dornheim_permutation_2015,militzer_development_2015,larkin_phase_2017,groth_configuration_2017,dornheim_ab_2018,dornheim_fermion_2019,yilmaz_restricted_2020,chang_recent_2015,church_real-time_2021,liu_unveiling_2020,vanbenschoten_piecewise_2022,zhang_finite_1999,he_reaching_2019} and deterministic methods\cite{kou_finite-temperature_2014,qin_finite-temperature_2021}, which treat the full space of excitations within the system.

Access to the electronic specific heat capacity and electronic entropy is in some cases straightforward; in other cases, more work is required. Within the low-cost methods, quantities can be calculated directly so long as not prohibited by memory requirements.\cite{kresse_efficient_1996,neugebauer_density_2013,khara_influence_2016,pourovskii_dynamical_2007}
As one increases the cost of the method, the computational cost quickly overruns practical access to calculating the specific heat capacity and entropy directly, and so one reverts to using numerical methods applied to the resulting finite temperature energy data.
Indirect access to the specific heat capacity and entropy has been explored in solid hydrogen chains using finite temperature auxiliary field quantum Monte Carlo (ft-AFQMC).\cite{liu_unveiling_2020} In this work, the researchers used an interpolation method to fit their finite temperature electronic energies as a function of temperature, thereafter calculating the specific heat capacity and entropy.
Another quantum Monte Carlo method has been used to calculate the electronic entropy, the interaction picture density matrix quantum Monte Carlo (IP-DMQMC) method. This was then used to calculate the free energy for the uniform electron gas.\cite{malone_accurate_2016}
The original density matrix quantum Monte Carlo method (DMQMC) and our recently developed piecewise interaction picture density matrix quantum Monte Carlo (PIP-DMQMC) have the benefit that almost every time step in the calculation simulates a new inverse temperature and the inverse temperature range is sampled very finely.
This provides a large amount of data for interpolation methods calculating the specific heat capacity and entropy, with the potential drawback of the stochastic noise in the energy potentially exceeding the change of that energy with changing inverse temperature.

Even with convenient access to a large data set, finite differences and cubic splines are susceptible to errors due to the underlying noise within the data set.
Machine learning has proven to be well-suited to quantum chemistry problems where there is an underlying relationship between an input variable and the resulting dependent variable.\cite{cheng_universal_2019,welborn_transferability_2018,bogojeski_quantum_2020,liu_toward_2022,lu_fast_2022,fiedler_deep_2022,borda_gaussian_2021}
Therefore, the application of machine learning methods, such as Gaussian process regression (GPR), may be well-suited for such interpolation problems as those required to calculate the specific heat capacity and entropy. In fact, the gradient of a Gaussian process regression fit is a well-defined and often-sought machine learning feature that is relatively straightforward to access in modern machine learning libraries.\cite{gpy_2014}
Furthermore, GPR is used in a diverse range of fields to fit data, typically with observational noise, some examples being: astrophysics\cite{strub_bayesian_2022,lackey_surrogate_2019}, geoacoustic surveying\cite{michalopoulou_matched_2021,frederick_seabed_2022}, energy science\cite{rahat_data_2018,rogers_probabilistic_2020}, and life science\cite{bahg_gaussian_2020,boukouvalas_BGP_2018}.
In each of these examples, noisy data are being fit with Gaussian Process Regression.

Our aim in this work is to create GPR models of finite temperature electronic energies from DMQMC simulations, which are comparable in quality to the original DMQMC data.
To test the quality of the GPR models, we first develop a workflow that depends only on the results of supervised fitting to find an optimized model. 

Secondly, we test the quality of the GPR models by calculating the specific heat capacity, which is found using the gradient of the energy with respect to temperature. The specific heat capacity from the GPR-derived gradients is then compared to the exact specific heat capacity, the specific heat capacity from numerical finite differences, and the specific heat capacity predicted by the derivative of the cubic spline fits. Next, we calculate the entropy both exactly and numerically, and compare the two. As a final test of the methods using nine molecular benchmark systems, we make comparisons of the many-body specific heat capacity and entropy to two approximate methods' specific heat capacity and entropy.
We finish by using the GPR method to calculate the electronic specific heat capacity and entropy for two molecular systems which have Hilbert spaces larger than what can be treated with exact diagonalization.
These tests are performed for molecular benchmark systems that we investigated in previous work.\cite{vanbenschoten_piecewise_2022}
In this study, molecules represent a benchmark system for the eventual treatment of \textit{ab initio} solids, and in the latter the electronic specific heat capacity is a more physically meaningful quantity.

\section{Methods}

\subsection{Finite temperature electronic structure methods}

\subsubsection{Deterministic methods}

We define the exact density matrix in the canonical ensemble as the sum over the full configuration interaction (FCI) states:
\begin{equation}
\hat{\rho}(\beta)=\frac{1}{\sum_i e^{-\beta E_i}}\sum_{i}e^{-\beta E_i}| \Psi_i \rangle \langle \Psi_i |,
\label{eq:rho}
\end{equation}
where $\beta=1/(k_{\mathrm{B}}T)$, and the FCI states $\Psi_i$ are defined as
\begin{equation}
| \Psi_i \rangle = \sum_j C_j^{(i)} | D_j \rangle,
\label{eq:psi}
\end{equation}
which are obtained from diagonalization of the Hamiltonian written in a Slater determinant basis: 
\begin{equation}
H_{ij}=\langle D_i | \hat{H} | D_j \rangle.
\label{eq:hamil}
\end{equation}
In order to study a range of systems with differing electronic structures, we keep the basis set and electron number relatively small. 
Furthermore, we only consider a single Hamiltonian symmetry (the symmetry of the ground state) when performing the sums over states.
We note that for physical descriptions of real system the sums over states would typically be carried out over all Hamiltonian symmetries.

From here, the energy is calculated as a sum over FCI eigenstates:
\begin{equation}
E(\beta)=\frac{\sum_i E_i e^{-\beta E_i}}{Z(\beta)},
\label{eq:energy}
\end{equation}
where $Z(\beta)=\sum_i e^{-\beta E_i}$, and then the specific heat capacity, 
\begin{equation}
C_V(\beta)=-\beta^2 \left(\frac{\partial E}{\partial \beta}\right)_{N,V},
\label{eq:defCv}
\end{equation}
can be found directly as a sum over FCI eigenstates by taking the derivative of \refeq{eq:energy} using the quotient rule:
\begin{equation}
C_V(\beta)=-\beta^2 \left[\left(\frac{\sum_i E_i e^{-\beta E_i}}{Z(\beta)}\right)^2-\frac{\sum_i E_i^2 e^{-\beta E_i}}{Z(\beta)}\right].
\label{eq:Cv}
\end{equation}

Two other approximate deterministic methods are also considered here: a thermal sum-of-states over the HF Slater determinants (THF), and the Fermi--Dirac FD thermal sum-of-states using the same HF Slater determinants (FDHF).
Performing the sum-of-states over all possible symmetries in the grand canonical ensemble, the FDHF method is equivalent to the traditional sum-over-orbitals FD method.\cite{fetter_quantum_2003,hirata_finite-temperature_2021}
However, we work in the canonical ensemble, and consider the ground state symmetry only, for which there are no known sum-over-orbital expressions.\cite{hirata_finite-temperature_2021}

The THF energy is calculated by working with the diagonal elements of the FCI Hamiltonian, such that the density matrix is:
\begin{equation}
\hat{\rho}_{\mathrm{THF}}(\beta)=\frac{1}{\sum_i e^{-\beta H_{ii}}}\sum_{i}e^{-\beta H_{ii}} | D_i \rangle \langle D_i |,
\label{eq:rhoTHF}
\end{equation}
where $D_i$ are the HF states, and the energy is:
\begin{equation}
E_{\mathrm{THF}}(\beta)=\frac{\sum_i H_{ii} e^{-\beta H_{ii}}}{\sum_i e^{-\beta H_{ii}}}.
\label{eq:THF}
\end{equation}

Using the same basis of Slater determinants from THF, we calculate the FDHF density matrix as
\begin{equation}
\hat{\rho}_{\mathrm{FDHF}}(\beta)=\frac{1}{\sum_i e^{-\beta F_{ii}^{(0)}}}\sum_{i}e^{-\beta F_{ii}^{(0)}}| D_i \rangle \langle D_i |,
\label{eq:rhoFDHF}
\end{equation}
where $F^{(0)}$ is the Fermi--Dirac Hamiltonian
\begin{equation}
F_{ij}^{(0)}=\langle D_i | \hat{F} | D_j \rangle,
\end{equation}
and $\hat{F}$ is the ground state Fock operator meaning the exchange contribution is based on the zero temperature occupied orbitals.\cite{szabo_modern_1996,hirata_finite-temperature_2021}
Then the FDHF energy is:
\begin{equation}
E_{\mathrm{FDHF}}=\frac{\sum_i F_{ii}^{(0)} e^{-\beta F_{ii}^{(0)}}}{\sum_i e^{-\beta F_{ii}^{(0)}}}.
\end{equation}

In these equations, we write (and subsequently calculate) the FDHF Hamiltonian in the Slater determinant basis in order to maintain consistency with the symmetry convention used by the other calculations (and because the systems are small enough to do so).
The method used here corresponds to the zeroth order many-body perturbation theory (MBPT0) in the canonical ensemble.\cite{hirata_finite-temperature_2021}
The MBPT0 internal energy is known to disagree with the Hartree--Fock energy at zero temperature.\cite{hirata_finite-temperature_2021}
MBPT0 can be improved upon by adding higher order perturbations.\cite{hirata_finite-temperature_2021,white_comment_2018}

\subsubsection{Density matrix quantum Monte Carlo methods}

Here, we provide an outline of the density matrix quantum Monte Carlo (DMQMC)\cite{blunt_density-matrix_2014} and piecewise interaction picture DMQMC (PIP-DMQMC) methods.\cite{vanbenschoten_piecewise_2022} 
The mathematical notation used and derivations below follow the two references cited, for consistency. 

In DMQMC the $N$-electron density matrix $\hat{\rho}(\beta)=e^{-\beta \hat{H}}$ is written in a finite basis of Slater determinants
\begin{equation}
\hat{f}(\tau)=\sum_{ij}f_{ij}(\tau)|D_i\rangle \langle D_j|.
\label{eq:Nrho}
\end{equation}
In the above equation, we replaced $\hat{\rho}(\beta)$ with $\hat{f}(\tau)$ for generality and consistency with other studies. The variable $\tau$ is the coordinate used during propagation of a DMQMC simulation. In the context of this equation it is a substitute variable equal to $\beta$, but this will not be the case for other methods.\cite{malone_interaction_2015,vanbenschoten_piecewise_2022}

The main objective in DMQMC methods is to calculate the energy,
\begin{equation}
E(\beta)=\frac{\Tr[\hat{f}(\beta)\hat{H}]}{\Tr[\hat{f}(\beta)]},
\label{eq:dmqmc_energy}
\end{equation}
by averaging the numerator and denominator calculated over $N_\beta$ simulations, individually referred to as $\beta$-loops.
To achieve this, the density matrix is sampled by evolving a population of signed walkers which reside on the sites $|D_i\rangle \langle D_j|$ during the simulation. 

Each DMQMC simulation begins with the known initial condition
\begin{equation}
\hat{f}(\tau=0)=\mathbb{1}.
\label{dmqmc0}
\end{equation}
The initial condition is sampled by randomly distributing walkers along the diagonal sites $|D_i\rangle \langle D_i|$ of the density matrix with a uniform probability.
Thereafter, the symmetrized Bloch equation,
\begin{equation}
\frac{\mathrm{d}}{\mathrm{d}\tau}\hat{f}(\tau)=-\frac{1}{2}\left[\hat{H} \hat{f}(\tau)+\hat{f}(\tau)\hat{H}\right],
\label{eq:bloch}
\end{equation}
is discretized using a finite time step
\begin{equation}
\Delta f_{ij}(\tau)=-\frac{\Delta\tau}{2}\sum_{k}H_{ik}f_{kj}(\tau)-\frac{\Delta\tau}{2}\sum_{k}f_{ik}H_{kj}.
\label{eq:bloch2}
\end{equation}
This is then combined with $f_{ij}(\tau)$, resulting in the equation of motion:
\begin{align}
f_{ij}(\tau+\Delta\tau)=
& \ f_{ij}(\tau)(1-\Delta\tau S)+\Delta f_{ij}(\tau)\notag\\
=&-\frac{\Delta\tau}{2}\sum_{k\ne i}H_{ik}f_{kj}(\tau)-\frac{\Delta\tau}{2}\sum_{k\ne j}f_{ik}H_{kj}\notag\\
&+f_{ij}(\tau)(1+\frac{\Delta\tau}{2}\left[2S-H_{ii}-H_{jj}\right]),
\label{eq:bloch_eom}
\end{align}
where $S$ is a constant energy shift used to control the total number of walkers during a simulation.
Application of this equation is broken into three steps: spawning, cloning, and annihilation.
These steps are based on the ground state method full configuration interaction quantum Monte Carlo (FCIQMC), and are outlined in Appendix \ref{appendix:dmqmc_algo}.\cite{booth_fermion_2009}
This similarity between the two methods means that DMQMC can potentially re-use many of the past and continuously developed methods for FCIQMC, such as excitation generators\cite{booth_fermion_2009,blunt_density-matrix_2014,neufeld_exciting_2018} and semi-stochastic adaptations.\cite{petruzielo_semistochastic_2012,blunt_semi_2015}
After several applications of the equation of motion, the energy terms are calculated and the shift is updated by:
\begin{equation}
S(\tau+\Delta\tau)=S(\tau)+\frac{\zeta}{A\Delta\tau}\ln{\left(\frac{N_w(\tau+\Delta\tau)}{N_w(\tau)}\right)},
\label{eq:shift}
\end{equation}
where $A$ is the number of steps between updates, $N_w$ is the total walker population for a given $\tau$, calculated as $N_{w}(\tau)=\sum_{ij}|f_{ij}(\tau)|$, and $\zeta$ is a damping parameter. The shift is only updated if a constant walker population is desired; it otherwise remains fixed.
The shift update equation is adopted from diffusion Monte Carlo\cite{umrigar_diffusion_1993}, motivated by the algorithm's success within FCIQMC.\cite{booth_fermion_2009}
Over the past couple of years there has been work to improve and develop new shift updating algorithms for FCIQMC which may also be used with DMQMC.\cite{yang_improved_2020,ghanem_population_2021,brand_stochastic_2022}
One example that may be useful is the shift update algorithm of Yang \textit{et. al.}, where the shift is approximated as a noisy damped harmonic oscillator.\cite{yang_improved_2020}

In interaction picture DMQMC (IP-DMQMC)\cite{malone_interaction_2015}, the initial density matrix is modified to be a trial density matrix that is then propagated to a target inverse temperature over the course of a simulation. 
This allows for a more accurate sampling of the initial condition for a large matrix, but has the disadvantage of only sampling one inverse temperature point. 
The PIP-DMQMC method\cite{vanbenschoten_piecewise_2022} extends the IP-DMQMC method to sample the density matrix for a range of temperatures rather than a single temperature, thereby reducing the computational cost of IP-DMQMC while retaining the desirable traits of IP-DMQMC.

In PIP-DMQMC, the simulation is conducted in two phases. In the first phase of the simulation, instead of sampling the density matrix, $\hat{\rho}(\tau)=e^{-\tau \hat{H}}$, the matrix sampled is
\begin{equation}
\hat{f}(\tau)=e^{-(\beta_T-\tau) \hat{H}^{(0)}}e^{-\tau \hat{H}},
\label{eq:aux}
\end{equation}
where $\beta_T$ is referred to as the target $\beta$. This matrix is only equivalent to the density matrix $f(\tau)=\rho(\tau)$ when $\tau=\beta_T$, which is also the point where the second phase of the simulation begins. In the second phase of the simulation, the Bloch equation (\refeq{eq:bloch}) is used to sample $f(\tau)$ for $\tau>\beta_T$.
The propagator for PIP-DMQMC is
\begin{align}
\frac{\mathrm{d}}{\mathrm{d}\tau}\hat{f}(\tau)=
&-\Theta(\beta_T-\tau)\left[-\hat{H}^{(0)}\hat{f}(\tau)+\hat{f}(\tau)\hat{H}\right]\notag\\
&-\frac{1}{2}(1-\Theta(\beta_T-\tau)\left[\hat{H}\hat{f}(\tau)+\hat{f}(\tau)\hat{H}\right],
\label{eq:pip}
\end{align}
where $\Theta(x)$ is the Heaviside step function and is equal to 1 when $x>0$ and otherwise equal to 0.

To begin the first phase, the initial condition is an approximate density matrix,
\begin{equation}
\hat{f}(\tau=0)=e^{-\beta_T \hat{H}^{(0)}},
\label{eq:pip0}
\end{equation}
where $\hat{H}^{(0)}$ is diagonal of the FCI Hamiltonian, i.e. $H_{ii}^{(0)}=H_{ii}$, and so the density matrix is equivalent to sampling the THF density matrix (\refeq{eq:rhoTHF}).
To stochastically sample the initial condition, we use the IP-DMQMC initialization protocol originally introduced by Malone and coworkers.\cite{malone_interaction_2015} The process generates determinants by attempting to occupy orbitals with a probability
\begin{equation}
p(i)=\frac{1}{e^{\beta_T(\epsilon_i-\mu)}+1},
\label{eq:fermi}
\end{equation}
where $\epsilon_i$ are the HF orbital energies and $\mu$ is a chemical potential that ensures the system's particle number $N$ is conserved. Walkers are then added to the site $|D_i\rangle \langle D_i|$ with a weight $e^{-\beta_T( \sum_i \epsilon_i)}$.

However, as pointed out by Malone and coworkers,\cite{malone_interaction_2015} this results in sampling an approximate density matrix for the grand canonical ensemble. Furthermore, this will generate the approximate density matrix for the Hamiltonian $\hat{H}^\prime\ne \hat{H}^{(0)}$. Malone and coworkers resolved this by first discarding those determinants which do not conserve the particle number $N$, as well as determinants outside the desired symmetry, resulting in the canonical density matrix $\hat{f}^\prime(\beta_T)=e^{-\beta_T \hat{H}^\prime}$. Then, to sample the density matrix for $\hat{H}^{(0)}$, the density matrix elements are multiplied by a reweighting factor:
\begin{equation}
f_{ii}(\beta_T)=f_{ii}^\prime(\beta_T) e^{-\beta_T (H_{ii}^{(0)} - H_{ii}^\prime)}.
\label{eq:reweight}
\end{equation}

The initial condition in PIP-DMQMC is then propagated using a finite time step as:
\begin{align}
\Delta & f_{ij}(\tau)=\notag\\
&\begin{cases}
-\Delta\tau\sum_{k}[-H^{(0)}_{ik}f_{kj}(\tau)+f_{ik}(\tau)H_{kj}], & \tau < \beta_T, \\
-\frac{\Delta\tau}{2}\sum_{k}[H_{ik}f_{kj}(\tau)+f_{ik}(\tau)H_{kj}], & \tau \ge \beta_T.
\end{cases}
\end{align}
When $\tau<\beta_T$, the same \textit{spawning}, \textit{cloning}, and \textit{annihilation} steps are used as in DMQMC, with the only changes being: \textit{spawning} now occurs only between $f_{ij}$ and $f_{ik}$, the factor $\frac{\Delta\tau}{2}$ becomes $\Delta\tau$, and the \textit{cloning} term is now $p_c=f_{ij}(\tau)\Delta\tau(S+H_{kk}-H_{jj})$. 

When $\tau\ge\beta_T$, the same steps are used as described above for DMQMC when using the symmetric Bloch equation (\refeq{eq:bloch}). However, another version of PIP-DMQMC used in this work uses the asymmetric Bloch equation (\refeq{eq:asym_bloch}) when $\tau\ge\beta_T$. Then the only modifications required are those described for \refeq{eq:asym_bloch}.

Like DMQMC, the main goal of PIP-DMQMC is to calculate the finite temperature energy using \refeq{eq:energy}.
However, unlike DMQMC, the finite temperature energy is only sampled for $\tau\ge\beta_T$.
The advantage of PIP-DMQMC is that it has been shown to reduce sampling issues present at small-to-intermediate $\tau$, typically referred to as `shouldering' in the energy estimate.\cite{vanbenschoten_piecewise_2022}
This is possible because the first phase of the simulation uses the interaction picture, which was originally developed to address the shouldering.\cite{blunt_density-matrix_2014,malone_interaction_2015}
The shouldering arises in some systems where there are many determinants, but the energy is largely dominated by a small set of low energy determinants. In these cases, the original DMQMC method’s initial condition ($\hat{\rho}(\tau\!=\!0)\!=\!\mathbb{1}$) can lead to sampling issues. This is because the diagonal of the density matrix is uniformly sampled, but not all rows are equally important to the simulation. This can be overcome in the original DMQMC method by increasing the number of opportunities to randomly select an important determinant by increasing the number of walkers or $\beta$-loops but doing so is generally undesirable. Instead, the IP-DMQMC method overcomes the sampling issue by starting the simulation with the thermal HF density matrix at the same temperature as some target temperature ($\beta_T$). This greatly increases the probability that important low energy states are present in the simulation from the start, thereby overcoming the sampling issue.\cite{malone_interaction_2015,petras_using_2020}
However, IP-DMQMC introduces a greater computational cost, because only one inverse temperature is sampled during a calculation.
The benefit of using PIP-DMQMC, is that data may be continuously collected for a range of temperatures beyond $\beta_T$ while simultaneously reducing the aforementioned shouldering.

The DMQMC methods suffer a sign problem due to the occurrence of positive and negative matrix elements in the density matrix.\cite{blunt_density-matrix_2014}
The sign problem in DMQMC is related to full configuration interaction quantum Monte Carlo (FCIQMC). The sign problem in FCIQMC had been systematically studied in the past\cite{booth_fermion_2009,spencer_sign_2012}, whereas the sign problem in DMQMC has only recently been systematically studied by several of us.\cite{petras_sign_2021}
Additional details on the sign problem and using the initiator approximation to overcome it can be found in Appendix \ref{appendix:initiator}.

\subsection{Taking energy derivatives}

\subsubsection{Finite differences}

One way to numerically approximate the gradient of a function is to use a finite difference method.
Additional details on the finite difference method, its benefits and shortcomings can be found in Appendix \ref{appendix:finite_difference}.

The protocol we used to compute a finite difference gradient is as follows.
DMQMC data are generated by averaging multiple independent simulations together using the post-analysis scripts provided in the PyHANDE package contained within HANDE-QMC\cite{spencer_hande-qmc_2019}. 
From the DMQMC dataset, we downsampled the DMQMC data, taking every tenth datapoint from the original dataset.
{\tt{NumPy}} was used to compute the numerical derivative using the gradient method.\cite{harris_array_2020}
The resulting gradient is multiplied by the negative of the inverse temperature squared to give the finite difference specific heat capacity.
We used the {\tt{NumPy}} library because we found it reasonable for our needs and because it is well documented.
This choice limited our ability to use the noise associated with the numerical data from DMQMC because the noise was not a parameter which could be supplied to the {\tt{NumPy}} function we used.
This limitation will not be investigated here, but we plan to return to considerations of noise in future investigations.

\subsubsection{Cubic Splines}

Another way to approximate the gradient of a function is to generate a cubic spline of the numerical data.
Cubic splines fit numerical data to a piecewise function, consisting of many third degree polynomials corresponding to subdomains for the independent variable of the original numerical dataset.
Additional details on cubic splines can be found in Appendix \ref{appendix:cubic_spline}.

Cubic splines were applied to the DMQMC data that had already been averaged across different beta loops, in the same way as finite differences. The {\tt{SciPy}} library was used to perform cubic spline generation and subsequent derivative calculations.\cite{mckinney_proc-scipy_2010,virtanen_SciPy_2020}
We used the {\tt{SciPy}} library cubic spline function because it was reasonable for our needs and the functionality was well documented.
As before with finite differences, this choice limited our ability to use the noise associated with the numerical data from DMQMC because the noise was not a parameter which could be supplied to the {\tt{SciPy}} function we used.
This limitation will not be investigated here, but we plan to return to considerations of noise in future investigations.
In order to improve some numerical noise, we found that we had to use a resampling interval, $N_\textrm{resample}$, which removes data points such that one data point per $N_\textrm{resample}$ is retained. 
In this implementation, the data in the interval was discarded. 
We settled on $N_\textrm{resample}=10$, and a full investigation into this choice can be found in Appendix \ref{appendix:cubic_spline}.

\subsubsection{Gaussian process regression}

We use GPy\cite{gpy_2014}, a Gaussian process framework in the Python programming language, because it supports various noise models and was recently successful in predicting quantum chemistry energies.\cite{welborn_transferability_2018,cheng_universal_2019} GPy additionally provides a method for attaining the predictive gradients which allow us to evaluate the specific heat (\refeq{eq:defCv}) as a function of temperature from fits to our DMQMC data.
The GPR kernel has two hyperparameters: a length scale, which determines how quickly the dependent variable varies with the independent variable, and a variance, which determines how far the predicting function can deviate from the mean. As the kernel parameters are optimized, GPR fits can be sensitive to the numerical challenges of optimization procedures. To help to avoid numerical issues in our fits in the high temperature limit, we separately fit the high-temperature DMQMC data as a function of inverse temperature ($\beta$) and the low temperature DMQMC data as a function of direct temperature ($T$). We then combined the $E(\beta)$ and $E(T)$ fits to form our final fit on the full temperature domain.
We used the same kernel for both the $E(\beta)$ and $E(T)$ fits for simplicity.
This kernel is comprised of a sum of three kernels: a radial basis function (RBF) kernel, a Matern 5/2 kernel, and a Matern 3/2 kernel.
We expect that this combination of kernels provides us with the flexibility to capture the exponential and polynomial behavior in the energies as a function of both $\beta$ and $T$.
The individual kernels are initialized with default parameters (a length scale and variance of 1.0) and optimized internally within GPy.
Finally, we note that GPR has been highlighted (Ref. \onlinecite{rasmussen_gaussian_2006}) as a suitable alternative to cubic splines, which can be justified by statistical arguments.

\begin{figure}
\includegraphics[width=0.45\textwidth,height=\textheight,keepaspectratio]{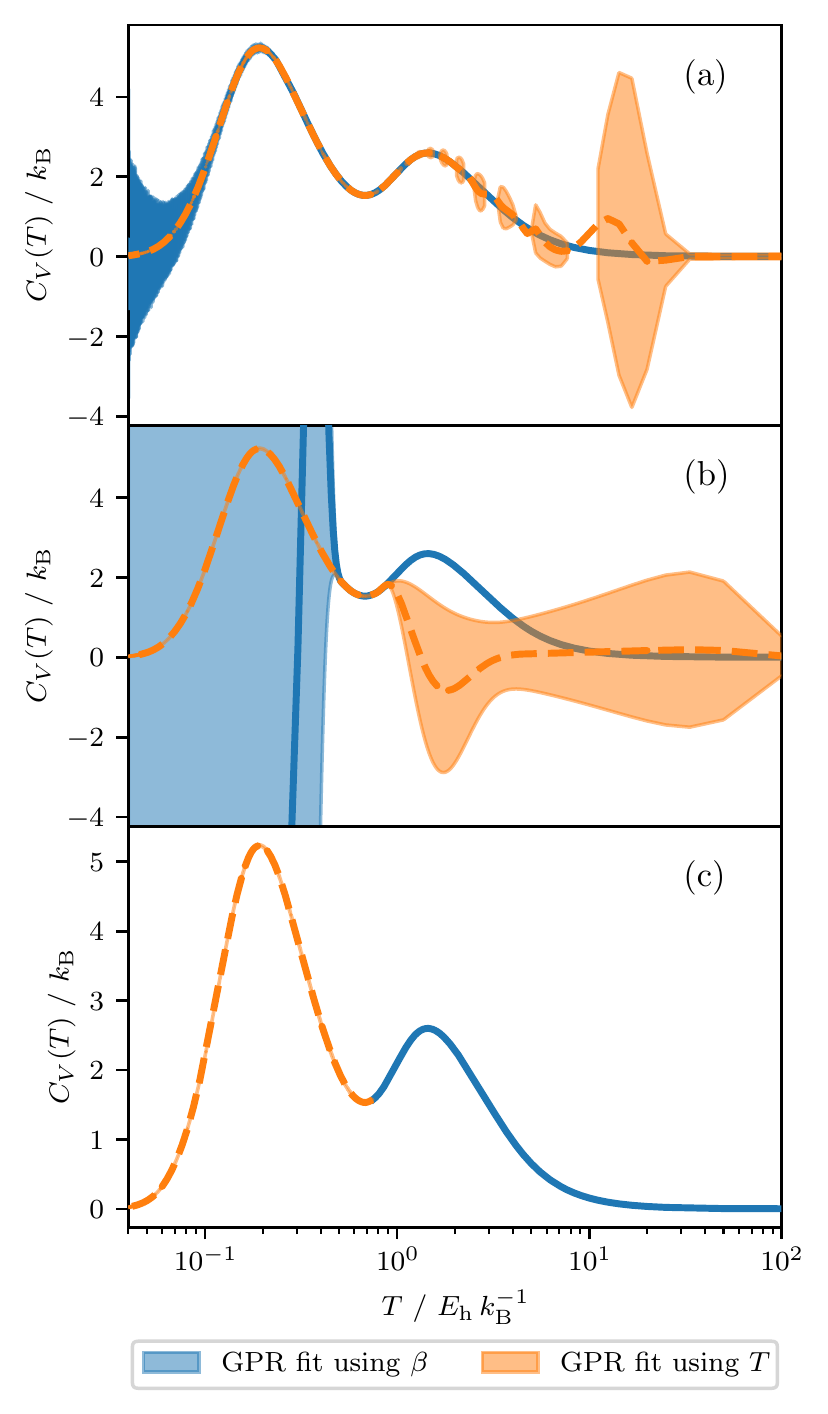}
\caption{An example of the three step process we use for calculating the specific heat capacity as a function of temperature using DMQMC data for the BeH$_2$ system.
(a) Initially, we fit using energy data from $\beta\in\left[0.0,50.0\right]$, which predicts the blue curve; 
and fit energy data from $1/\beta=T\in\left[0.0,10.0\right]$, which predicts the orange curve.
(b) Then, the data is refit on $\beta\in\left[0.0,2.0\right]$ (blue) and $T\in\left[0.0,1.0\right)$ (orange).
(c) The final step is to combine the fits using the blue curve for $T\ge0.75$ and the orange curve for $T<0.75$.}
\label{fig:model}
\end{figure}

The initial GPR fits were performed on downsampled DMQMC and PIP-DMQMC data, where every tenth data point was used instead of the entire data set. 
Then, predictions were performed on the original $\beta$- (or $T$-) grid.
This downsampling was chosen for efficiency, because we achieved no benefit in using the whole data set. 
The only exception to this were the $\beta$ fits for N$_2$/STO-3G and LiF/STO-3G, which used the entire dataset.
Unless indicated otherwise, the fits were done using the average energy estimate from DMQMC and PIP-DMQMC.
For GPR fits using $T$ as the independent variable, the ground state energy from FCI or FCIQMC was added to the training data using $T=0$ as the corresponding independent variable. When $\beta$ was used as the independent variable, the ground state energy was added using $\beta=50$ as corresponding independent variable. Once the ground state is included, the initial GPR models were trained using the down sampled data with the included ground state.

\subsubsection{Treatment of noise}

Throughout the manuscript we will use interpolation and gradient methods which do not require (or use) the variance associated with a given set of numerical data.
The choice to not use the variance associated with a numerical data is primarily due to the literature precedence.
To our knowledge the cubic spline and homoscedastic gaussian process, both of which do not use the variance for a given data set, are frequently used within the electronic structure community.\cite{liu_unveiling_2020,welborn_transferability_2018,kananenka_efficient_2016,von_Lilienfeld_retrospective_2020,motta_towards_2017}
These methods are what we used here.

There are alternatives to both the cubic spline and homoscedastic gaussian process which can incorporate the variance associated with a data set to improve interpolation estimates.
For spline methods, the univariate spline methodology can incorporate the variance as weights to improve the final piecewise polynomial fit.\cite{heuer_integrated_2018}
In the case of GPR, a heteroscedastic noise model may be used.\cite{maier_bypoassing_2022}

\section{Calculation details}

\subsection{Integrals}

The systems used in this work are entirely defined by the one- and two-particle interactions between the electrons, the electrons and the nuclei, and in some cases an additional core Hamiltonian term.
To generate the integrals defining our systems, we first run a restricted HF calculation, and the resulting orbitals are used to generate all integrals and the core Hamiltonian term. These terms are dumped into a text file, referred to as an FCIDUMP, and are organized by orbital indices. 
The Molpro quantum chemistry package was used to perform these calculations.\cite{werner_molpro_2019}

The HANDE-QMC software additionally requires orbital energies to be included in the FCIDUMP file. The Molpro package provides access to these; however, to achieve our desired level of accuracy, we found using a Python script to add the orbital energies to each FCIDUMP was more appropriate.
We validated the use of DMQMC to study molecular Hamiltonians in a previous study.\cite{petras_using_2020} 
Finally, we note our calculations use a finite basis set, and as such will contain basis set incompleteness errors.
The functional form of basis set error has been well studied for solids at finite temperature.\cite{malone_interaction_2015}
However we are unaware of similar studies for molecules at finite temperature, and as such we are unsure of the functional form for the basis set error and whether extrapolations to remove basis set error are feasible due to numerical issues resulting from Rydberg states.\cite{strickler_electronic_1966}

\subsection{Data acquisition}
\label{sec:calc}

Most of the DMQMC and all of the PIP-DMQMC data, as well as the eigenstates for calculating sum-over-states quantities, were collected using the open source HANDE-QMC software package, as part of a previous investigation conducted by several of the authors of this work.\cite{spencer_hande-qmc_2019,vanbenschoten_piecewise_2022}
Data from our previous investigation, used in our current work, was accessed using the same files available in the relevant public data repository.\cite{shepherd_dataset_2022} 

The system and basis set combinations used in this investigation are: Be/aug-cc-pVDZ, BeH$_2$; Be/cc-pVDZ; H/DZ, equilibrium H$_4$/cc-pVDZ, CH$_4$/cc-pVDZ, H$_2$O/cc-pVDZ, and STO-3G for CO, HCN, LiF, N$_2$, equilibrium H$_8$, stretched H$_6$ and stretched H$_8$. 
The geometries associated with these systems are available within the supplemental material for our previous investigation.\cite{vanbenschoten_piecewise_2022} The H--H bond distance for stretched H$_6$ is the same as stretched H$_8$.

The sum-over-states quantities for ft-FCI and THF are calculated using the eigenstates of the FCI Hamiltonian and the diagonal elements of the FCI Hamiltonian, respectively. The eigenstates were read in from HANDE-QMC output files for the benchmark systems. The sum-over-states calculations were performed using in-house Python scripts. The Fermi--Dirac (FDHF) eigenstates were calculated by first generating all the allowed occupation vectors using {\tt{SymPy}} functions\cite{meurer_sympy_2017}; thereafter, the entire set of occupation vectors has its corresponding energy calculated using the appropriate formula. Similarly, the THF eigenstates for water and methane are generated using the same occupation vectors from FDHF along with the Slater--Condon rules.\cite{szabo_modern_1996}
This entire process was performed using in-house Python scripts that read in only the integral file for each system. The full array of eigenstates for all occupation vectors was used in the same fashion as the ft-FCI and THF eigenstates to calculate the FDHF thermodynamic quantities.

For systems which have analytical ft-FCI results of the specific heat capacity, a total of $N_\beta=100$ were used to generate the energy estimates from DMQMC.
The DMQMC simulations were run with a system dependent target walker population.
The system and target walker populations were: Be/$N_w=5\times10^5$, BeH$_2$/$N_w=1\times10^7$, CO/$N_w=5\times10^5$, equilibrium H$_4$/$N_w=1\times10^7$, equilibrium H$_8$/$N_w=1\times10^7$, stretched H$_8$/$N_w=1\times10^7$, HCN/$N_w=1\times10^7$, LiF/$N_w=5\times10^6$, and N$_2$/$N_w=5\times10^5$.
The number of simulations and target walker populations for the PIP-DMQMC calculations for CH$_4$ was $N_\beta=5$/$N_w=5\times10^8$, and for H$_2$O was $N_\beta=4$/$N_w=5\times10^8$.
For the DMQMC calculations, the number of simulations and target walker populations was $N_\beta=5$/$N_w=1\times10^9$ for CH$_4$ and $N_\beta=4$/$N_w=1\times10^9$ for H$_2$O.
The walker population for the DMQMC for the CH$_4$ and H$_2$O calculations was increased to compensate for increased diagonal death rates in DMQMC.
For CH$_4$ and H$_2$O, DMQMC calculations were run from $\beta=0$ to $\beta=1$, and PIP-DMQMC simulations used a target $\beta_T$ of $\beta_T=1$ to obtain data for $\beta_T \ge 1$. This was done for consistency with the data from Ref.~\onlinecite{vanbenschoten_piecewise_2022} which we are using here.
Finally, the walker populations for all stretched H$_6$ DMQMC simulations was $N_w=1\times10^5$.

Unless indicated otherwise, all DMQMC and PIP-DMQMC simulations used a timestep of $\Delta\tau=0.001$, ten iterations between data reporting ($A=10$), a damping parameter of $0.05$ ($\zeta=0.05$), and the energy shift was always varied to keep the walker population approximately constant throughout the simulations. To calculate the average energy and error on the average energy for DMQMC and PIP-DMQMC data, the finite temperature analysis script provided within HANDE-QMC was used.

The DMQMC and PIP-DMQMC energy data for the finite difference, cubic spline and GPR methods was read in using the {\tt{pandas}} Python library.\cite{reback_pandas_2020,harris_array_2020} Data manipulations to prepare data for method application, and to perform additional analysis, were done using {\tt{pandas}} and {\tt{NumPy}}.
All Python libraries used in our calculations and analysis were acquired using the Conda installer, accessed through the open source Anaconda individual edition.\cite{anaconda}

\section{Results and discussion}

\subsection{Training and validating the GPR model} 

\begin{figure*}
\includegraphics[width=1.0\textwidth,height=0.5\textheight,keepaspectratio]{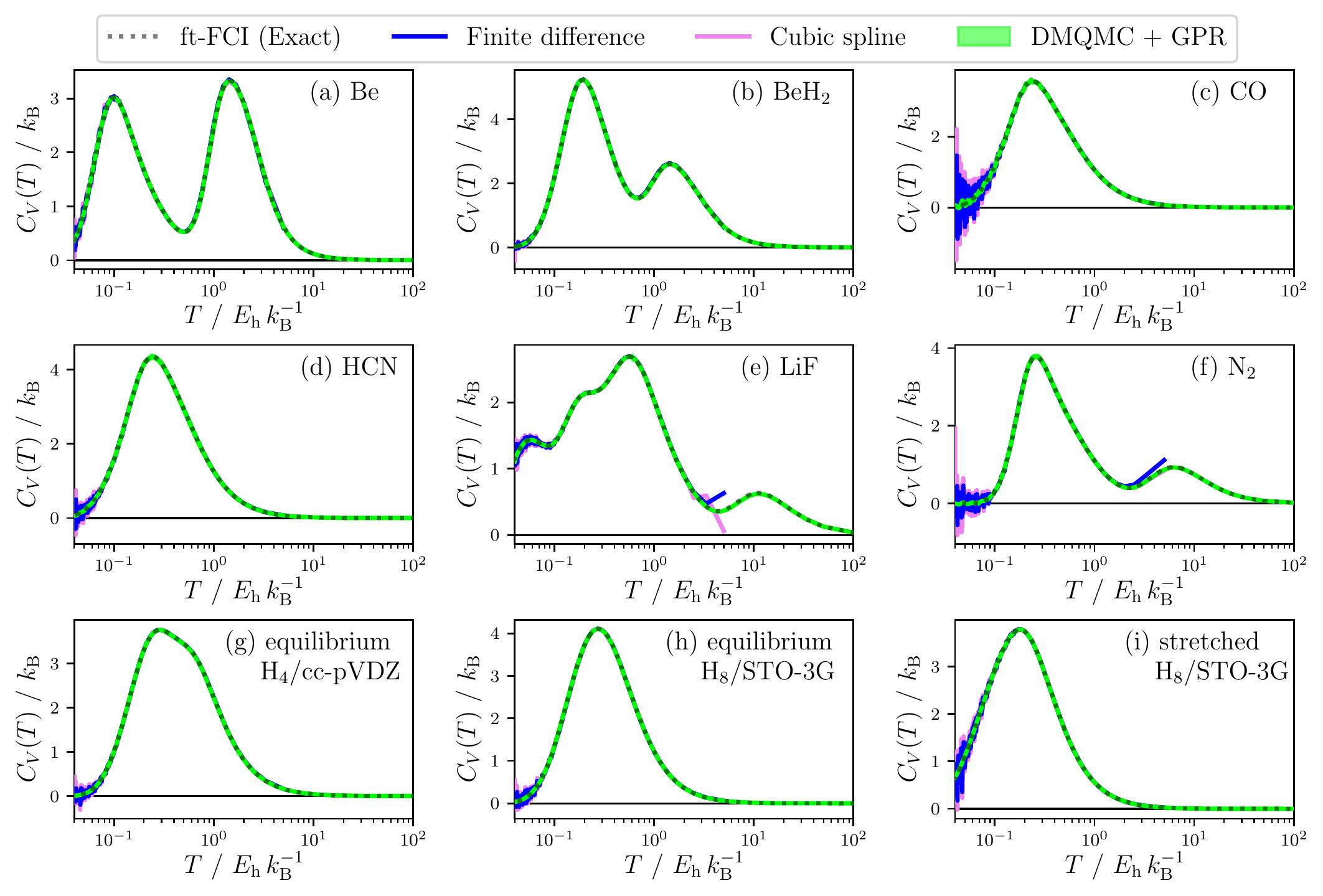}
\caption{Specific heat capacities ($C_V$) calculated as a function of temperature using several methods for (a)-(i) a range of benchmark systems whose Hamiltonians can be exactly diagonalized. Three methods of computing the $C_V$ from the DMQMC data are shown. 
The $x$-axis temperature window is fixed from $T=0.04$ to $T=100.0$, which corresponds to the physically-meaningful temperatures contained in the original DMQMC data set. Additionally, a solid black line is plotted at $C_V=0.0$ to aid with interpretation. The finite difference and cubic spline $C_V$ are only calculated/shown up to $T=5.0$ due to our downsampling procedure. For intermediate $T$, the finite difference and cubic spline $C_V$ is typically not visible due to overlap with GPR.}
\label{fig:Cv}
\end{figure*}

\begin{figure*}
\includegraphics[width=1.0\textwidth,height=\textheight,keepaspectratio]{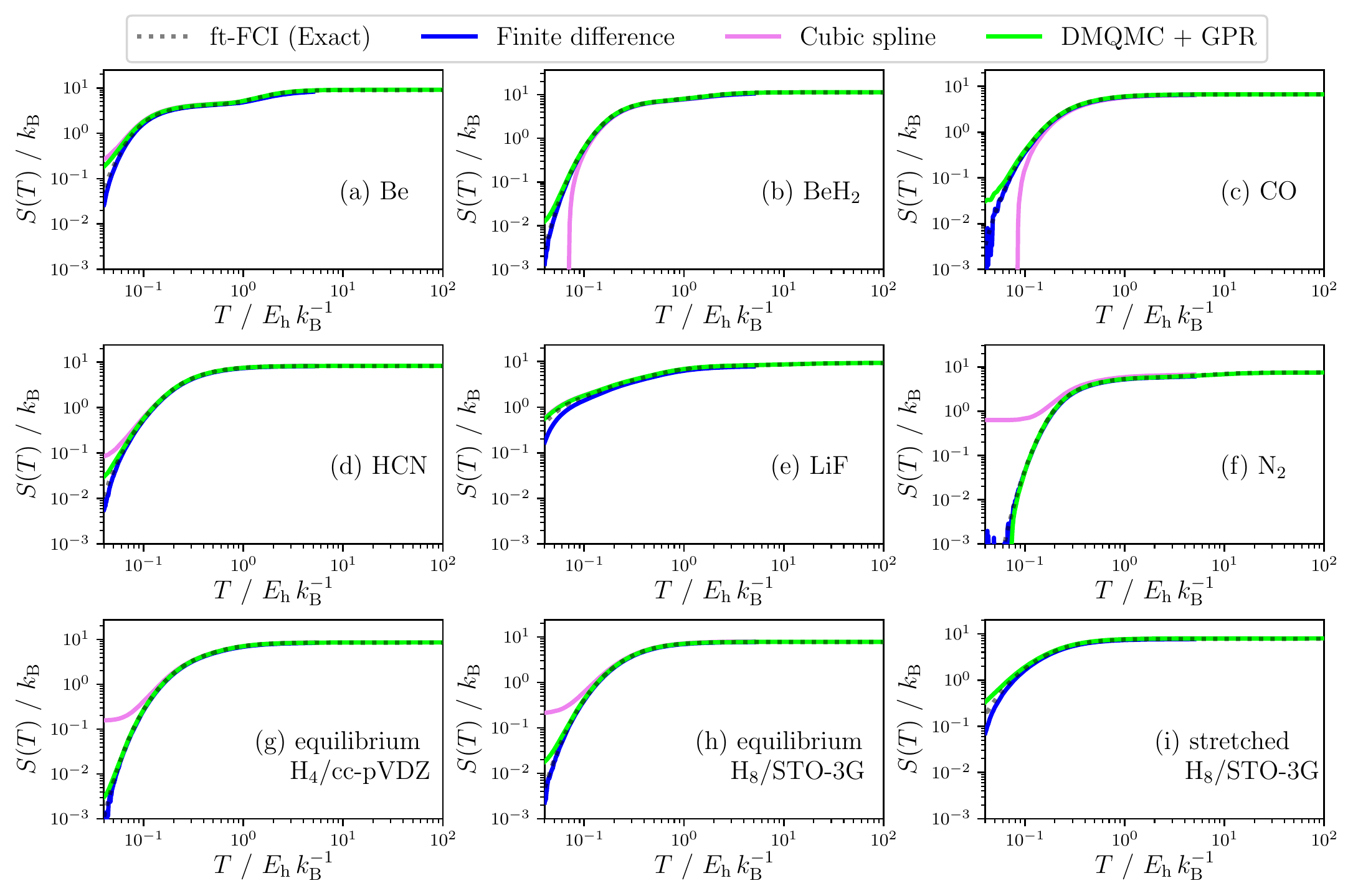}
\caption{The electronic entropy ($S$) calculated as a function of temperature using several methods for (a)-(i), which are the same benchmark systems shown in \reffig{fig:Cv}.
To calculate the exact $S$ (black dashed) we explicitly carried out the sum-over-states calculation in \refeq{eq:sosentropy}.
The electronic entropy is only shown up to $T=5.0$ for the finite difference method and up to $T=3.3$ for the cubic spline method due to the range limitation of the data set in \reffig{fig:Cv} (see text).}
\label{fig:entropy}
\end{figure*}

To optimize the GPR model, we used a three-step process. 
In the first step, we created two separate GPR models, one that takes in the energy data as a function of $\beta$, and another that takes in the energy data as a function of $T$. 
To these energy data, the ground-state FCI or FCIQMC energy was added at $\beta=50$ for the fit on $\beta$ (where we had data, we checked that this was large enough that the ftFCI energy was within 1mHa of the FCI energy) and $T=0$ for the fit on $T$.
Our motivation for using separate models was that they had different performance at high and low temperatures. 
Both domains have a similar appearance for intermediate domain values, and are expected to have similar performance in this region. 
At large $\beta$, the energy function becomes more slowly varying with $\beta$ and is more difficult for the GPR model to fit, as noise gradually increases. In this regime, the fit on $T$ tends to perform well. 
The converse was true at small $\beta$, where the fit on $T$ tended to become noisier. 
Once the two fits were generated, we calculated the specific heat capacity using the model's prediction of the energy gradient. The two specific heat capacities were then plotted, and visual inspection was used to determine subdomains where the fits are performing optimally by comparing the noise between either model's inverted counterpart.

For the second step, the GPR models were generated a second time for each domain. This time, only the subdomains where the models appeared least noisy were used, with the constraint that both models' subdomains overlapped with one another and, when combined, covered the entire domain of the original data set.
Once again, the specific heat capacity was calculated using the predictions of the energy gradient. The specific heat capacity was plotted and visually inspected to ensure the predictions generally remained consistent between the first step and the second step for the new subdomain. We then identified a cross-over between the two models' domains, which produced a minimally-noisy specific heat capacity curve. 

If the behavior drastically changed between steps one and two for the new domain, or a cross-over was not identified, the subdomains used in the second step were hand adjusted until the desired behavior was achieved.
For the final step, the two models were combined as a piecewise function and used to make predictions on the whole of the original domain.
We switched which model was used to make predictions at a crossover point, which was chosen by visual inspection.
The result was two subdomains without overlapping points, ensuring a single-valued function.
There was a slight jump between the two domains, but this did not seem to appreciably affect our heat capacities or entropies.
As such, we did not investigate a switching function.

To demonstrate the three-step process for creating a GPR model of the DMQMC energy data, we show the figures resulting from each step of generating a model for the BeH$_2$ system in \reffig{fig:model}. 
In \reffig{fig:model}(a), both models' predictions are shown using the entire domain for training. The model that uses $\beta$ as the domain, shown in blue, tends to have minimal noise in the high-to-mid temperature range. By contrast, the model that uses $T$ as the domain, shown in orange, tends to have minimal noise in the low-to-mid temperature range. Furthermore, we can see that there is an intermediate temperature regime where both models are of equal quality. 
Informed by \reffig{fig:model}(a), we guessed a crossover of $T~1.0$, and created updated models using training data from $T\in\left[0.5, \ 100.0\right]$ for the $\beta$ domain model, and $T\in\left[0.0, \ 1.0\right)$ for the $T$ domain model using the procedure described above. 
The new models' predictions for the entire domain are shown in \reffig{fig:model}(b). Inspecting the second panel, we can observe that both models are visually equivalent to the first panel for the new subdomains. Finally, in \reffig{fig:model}(c), we constructed the final curve using $T=0.75$ as the crossover. Minimal noise is observed across the entire temperature range.

The goal of this process was to find a prediction of the specific heat capacity that has minimal noise across temperature, which \reffig{fig:opt}(c) is found to achieve. The process of refitting, shown in \reffig{fig:opt}(b), is done to improve the GPR fits by only training on the ranges of relevant data, with some amount of overlap so both models connect smoothly.

To validate our GPR method,
we applied it to nine benchmark systems from the literature to calculate the specific heat capacity. 
These were then compared to the exact specific heat capacities, as well as those calculated from the other two numerical gradient methods presented here: cubic splines and finite differences. 
Figure \ref{fig:Cv} shows all nine systems in separate panels, with each panel showing the four methods overlaid.

For each system, we observe that all three approximate methods agree well with the exact data for the intermediate $T$ range. 
This observation is impressive, as numerical differentiation and fitting tends to be difficult for data from stochastic methods.
This is probably most remarkable for the LiH system, shown in panel \reffig{fig:Cv}(e), where an unanticipated stair-case-like pattern emerges as the temperature decreases to the ground state. 
The different methods do have a slight disagreement with the exact $C_V$ due to DMQMC having some `shouldering' in the energy at intermediate $\beta$ ($0.2<T<1$).
By way of an example, an energy error of $\sim 0.004$ Ha causes an error in the $C_V$ of $<0.01$ for equilibrium H$_4$/cc-pVDZ. This was the system with the largest shouldering error of this type.

GPR shows improvements over the cubic spline and finite difference approaches at both higher and lower $T$. 
When approaching the ground state ($T\rightarrow 0$), the GPR method controls noise much more effectively than the other two methods. 
The amount of noise at low $T$ does appear to be system dependent: CO, N$_2$, and stretched H$_8$ systems have the largest noise at low $T$. However systems like BeH$_2$, Be and equilibrium H$_4$ have minimal noise at low $T$. 
Regardless of the observed amount of noise due to the system, the GPR model is generally observed to have noise small enough that it is either difficult to observe, or cannot be seen on these scales due to the width of the line. 
A typical error for the GPR is $0.00001-0.03$, as seen in N$_2$ (\reffig{fig:Cv}(f)).

Improvements are also observed when using GPR for the high $T$ limit. 
In the high $T$ limit, the finite difference and cubic spline approaches stop providing data, due to having no way of producing a value at the edge of the data set. 
This is somewhat exacerbated by our use of every tenth data point from the original DMQMC data set, which was required to minimize the noise (and systematic error) in all three methods.
By contrast, GPR is able to reproduce the behavior of the exact specific heat capacity out to $T=100$.
For the majority of the systems in \reffig{fig:Cv}, the high-temperature specific heat capacity approaches zero, and the performance of all three methods is considered equivalent. 
Two notable outliers are LiF (\reffig{fig:Cv}(e)) and N$_2$ (\reffig{fig:Cv}(f)), where there is still significant non-zero behavior in the specific heat capacity at large temperatures that GPR is able to recover.
In both cases, the finite difference approach fails, and in LiF the cubic spline approach also has significant error.

\subsection{Calculating the electronic entropy} 

The entropy has been calculated in the interaction picture variant of DMQMC before by integrating the interacting part of the Hamiltonian over a trajectory in $\beta$,\cite{malone_accurate_2016} but this requires access to the IP-DMQMC trajectory, which has the drawback of calculating one $\beta$ at a time. 
Our method presented here provides another way to obtain the electronic entropy. 

We numerically integrated data in 
\reffig{fig:Cv} to calculate the electronic entropy as:
\begin{equation}
S(T)=\int_0^{T} C_{V}(T^\prime)~\frac{1}{T^\prime}~\mathrm{d}T^\prime,
\label{eq:definite}
\end{equation}
where $T$ is the temperature for the resulting entropy for the system.
These integrations were performed numerically using the trapezoid rule (as implemented in {\tt{NumPy}}); an entropy of zero was added into the data sets at zero temperature as required.
For the exact entropy, we instead used an analytical expression:
\begin{equation}
S=-k_\mathrm{B}\sum_i\left[\frac{e^{-\beta E_i}}{\sum_i e^{-\beta E_i}} ~ \ln{\left(\frac{e^{-\beta E_i}}{\sum_i e^{-\beta E_i}}\right)}\right],
\label{eq:sosentropy}
\end{equation}
where $E_i$ are the the FCI eigenvalues.

Figure \ref{fig:entropy} shows the resulting entropy calculations, where the same four methods are compared as for the previous section.
We note that the cubic splines entropy is shown up to $T=3.33$, the $T$ before $T=5.0$, which we removed based on our judgment that the $T=5.0$ entropy was nonphysical due to our methodologies' numerical limitations.

In the intermediate $T$ range, we observe that all three methods are able to produce reasonable entropy estimates. Where data is available for all three methods, there tends to be good overlap with the exact entropy, and generally, the three are similar until the curve begins to significantly change as the low-$T$ range is approached. 
The most obvious issue at low $T$ arises in the cubic spline estimates; these both over- and under-estimate the entropy. These errors can be dramatic, as in the case of CO (\reffig{fig:Cv}(c)) and N$_2$ (\reffig{fig:Cv}(f)). 
This is a result of the increased noise of the $C_V$ estimates, which are more pronounced for the cubic splines, and the loss of the edge data point from our data set. 

Though with very close inspection, some more subtle deviations from the exact entropy are seen for all methods at all temperatures. 
Some examples of this for GPR are the Be atom (\reffig{fig:entropy}(a)) and stretched H$_8$ (\reffig{fig:entropy}(i)), where GPR is seen to slightly overestimate the entropy. 
For the finite difference entropy, a consistent underestimation is observed, such as in the Be atom (\reffig{fig:entropy}(a)) and LiF (\reffig{fig:entropy}(e)). 
Underestimates of the cubic spline entropy is seen in BeH$_2$ (\reffig{fig:entropy}(b)), while overestimation of the entropy is seen in N$_2$ (\reffig{fig:entropy}(f)).

\begin{figure*}
\includegraphics[width=1.0\textwidth,height=\textheight,keepaspectratio]{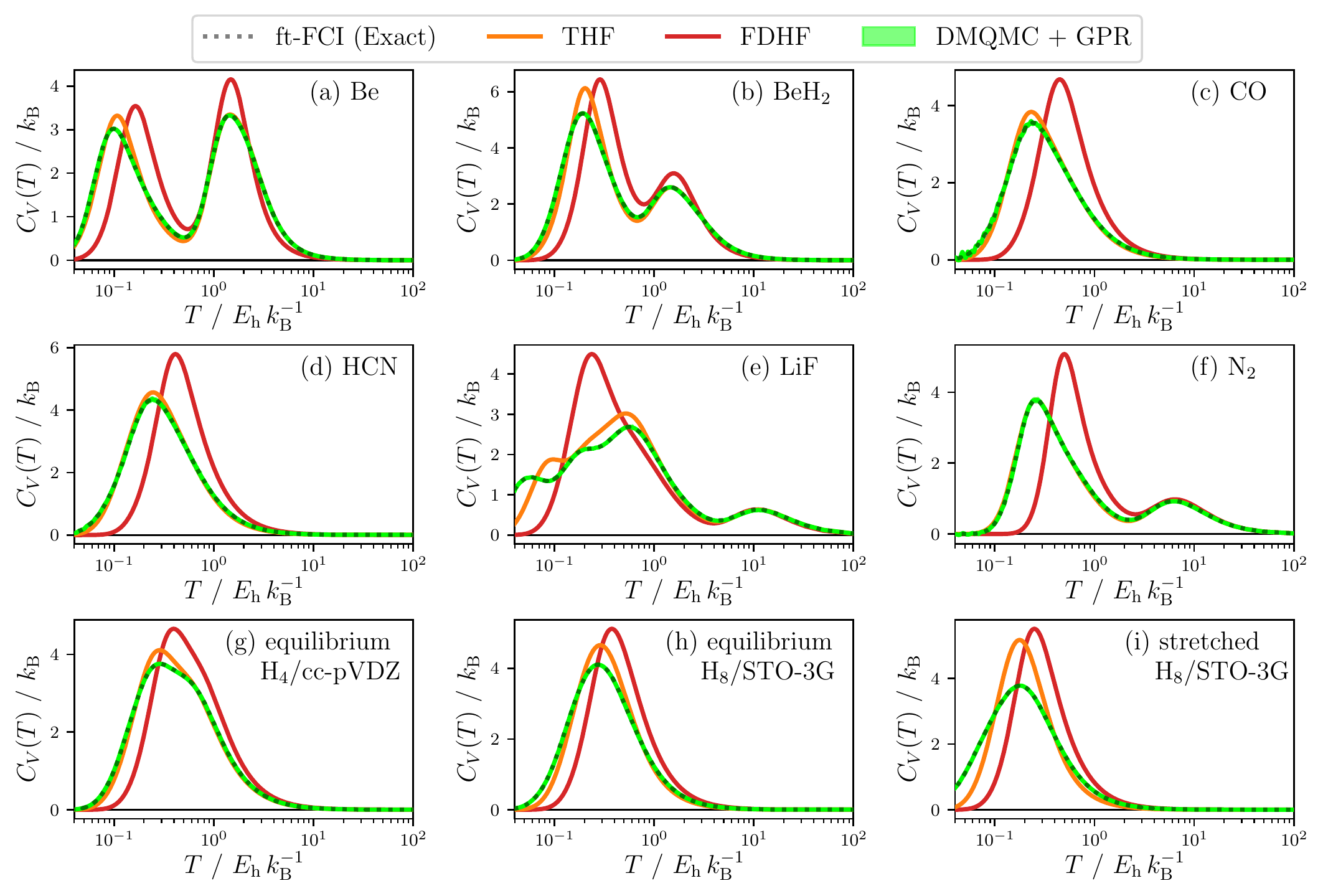}
\caption{The specific heat capacity from GPR fits of DMQMC energy data (lime) compared to two approximate methods, THF (orange) and FDHF (red). The panels (a)-(i) contain the same benchmark systems used in \reffig{fig:Cv} and \reffig{fig:entropy}.
For all figures, a solid black line is included at $C_V=0.0$ to aid in interpretation, and the temperature window is fixed from $T=0.04$ to $T=100.0$.
Data from GPR (lime) is the same data shown in \reffig{fig:Cv}.}
\label{fig:Cv2}
\end{figure*}

In general, all three methods behave consistently across temperatures when comparing their entropy estimates to the exact entropy.
Although all three methods can under- or overestimate the entropy, the consistent underestimating of the finite difference method, and nearly consistent overestimating of the GPR method across systems lends itself to a more consistent prediction, which is generally more desirable.
The deviations seen in these methods were also minor compared to the deviations seen in the cubic spline estimates.
The overshooting is related to the trapezoid rule. As we start the integration from $T=0$, the first step to get to the first $C_V$ point is linear in $T$ and tends to be an overestimate. An underestimate is caused by noise as it requires $C_V$ points to be lower than the exact $C_V$, causing a downward error in the entropy.

We also note that the GPR provided a larger range of entropy estimates (i.e., for the entire domain of $T$). 
This is because the finite difference and cubic spline methods are only able to be integrated and produce entropy estimates up to the limits of their specific heat capacity data in \reffig{fig:Cv}. 
Though this only has meaningful implications for those few systems, such as LiF (\reffig{fig:Cv}(e)), which had significant non-zero specific heat capacities at large $T$. Such a fact, though not detrimental in these circumstances, may be more important for different systems.
Thus, as the GPR method is able to produce entropy estimates for the entire range of temperatures, we consider it the optimal choice for calculations like the entropy in \reffig{fig:entropy}.

\subsection{Comparing the specific heat capacity and entropy between DMQMC, THF, and FDHF}

\begin{figure*}
\includegraphics[width=1.0\textwidth,height=\textheight,keepaspectratio]{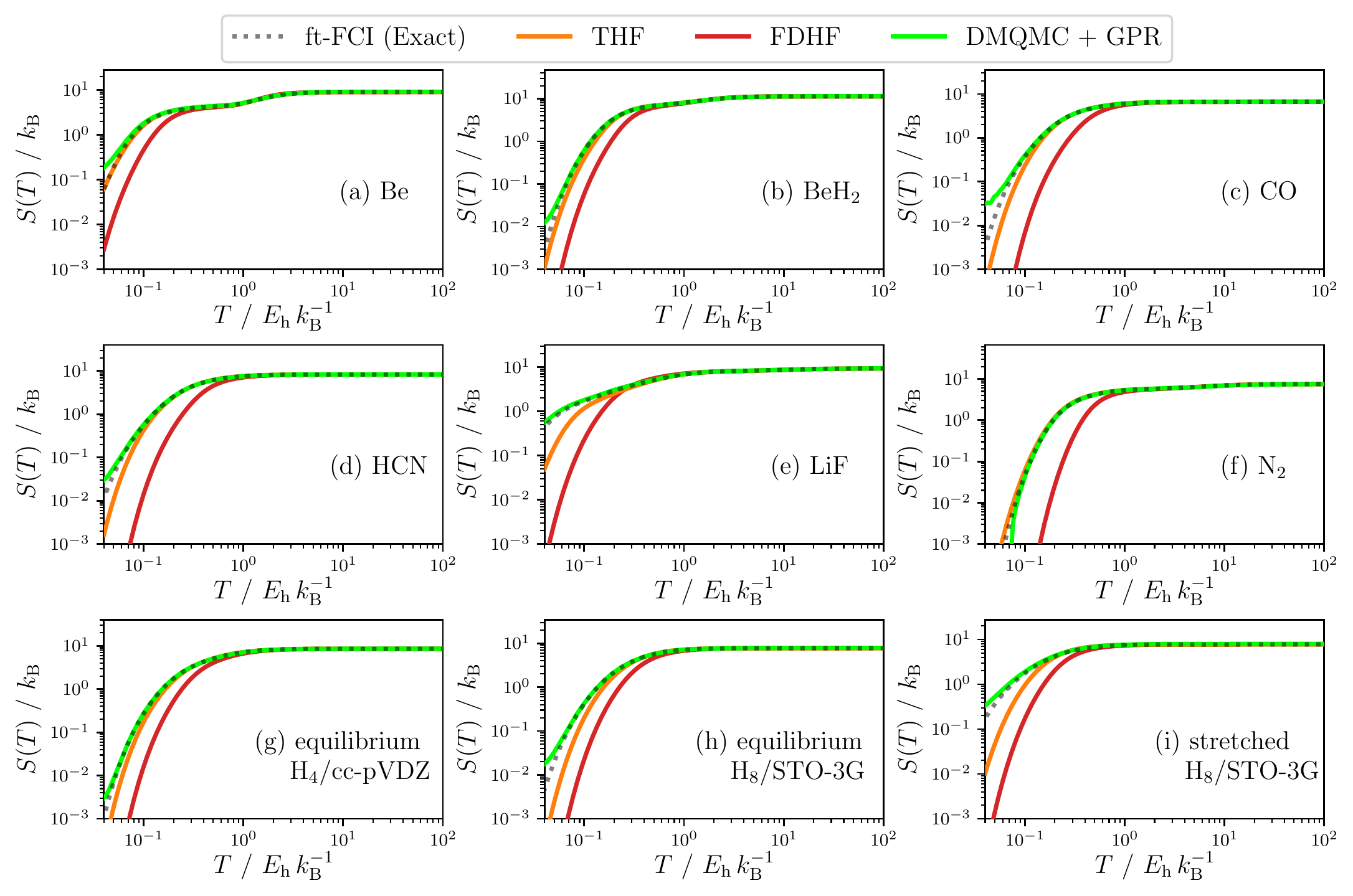}
\caption{The electronic entropy ($S$) calculated as a function of temperature is compared for DMQMC, THF, and FDHF methods. Panels (a)-(i) are the same benchmark systems shown in previous figures, and as such the DMQMC~+~GPR data are the same as in \reffig{fig:entropy}.}
\label{fig:entropy2}
\end{figure*}

Having validated the GPR method for calculations of the DMQMC specific heat capacity and the electronic entropy, we investigate the GPR method's ability to predict behavior relative to two other approximate methods. 
The first method, which we call THF (thermal HF), uses a Boltzmann-weighted ensemble of Slater determinants, i.e., the density matrix is constructed from the diagonal of the FCI matrix. 
The second method, Fermi--Dirac (FDHF), uses the same set of states as THF, but uses the ground state HF orbital energies to build the density matrix. We compute this estimator in this way for consistency, as we only sample the lowest-energy symmetry of the Hamiltonian. 
Figure \ref{fig:Cv2} shows the specific heat capacities for exact and DMQMC (from GPR) for the same set of benchmark systems used in \reffig{fig:Cv}, with the addition of the THF and FDHF specific heat capacities.oppu

In the high-$T$ limit, we expect each method to have a similar heat capacity, as our systems are simulated using a finite basis and particle number. This means there are a finite number of energy states available, and they become saturated as $T$ increase. When the energy states are saturated, the energy of the system can no longer change, and by \refeq{eq:Cv}, the specific heat goes to zero. This behavior, or a trend towards it, is observed for all methods. It can be noted that the approximate temperature for which the specific heat capacity approaches zero as predicted by each method appears to be similar for all systems, which is consistent with the high-temperature-limit density matrix being the identity matrix in any basis. 
It is important to note that our calculations here used a fixed geometry with a finite basis, which ignores any effect of continuum states accessed by electron ionization.
As such, the high temperature limit beyond a system's first ionization energy is not intended to be interpreted physically.
As an example, the experimentally measured first ionization energy for N$_2$ is $T\!\sim\!0.573$ Ha.\cite{trickl_state-selective_1989}
Furthermore, the electronic states of an isolated hydrogen like atom have modelling challenges when the basis set is completed, including a divergent entropy.\cite{strickler_electronic_1966} We have no evidence that molecules would be any different. However, for a finite basis set, thermodynamic quantities are well behaved. Temperature, here, is an external means of increasing the internal energy of our system by increasing the occupation of excited states in an equilibrium fashion. Specific heat capacity is the availability of the electronic energy levels of the molecule to undergo an energy change upon changing temperature.
Given these two facts, we are more focused on the investigating the consistency of each method as each temperature-accessible energy window accessible runs out of states analogous to what would happen at a band edge.

The specific heat capacity will also approach zero as $T$ approaches zero. This behavior is expected for all methods, and is a result of the energy becoming constant as each method settles to their respective ground states. We again observe this behavior, or a trend towards it, for all the methods presented.
The temperature at which the specific heat capacity is zero can be significantly different between methods for certain systems, which can be seen in LiF (\reffig{fig:Cv2}(e)) and N$_2$ (\reffig{fig:Cv2}(f)) when comparing the DMQMC and FDHF specific heat capacities.

Finally, the intermediate $T$ range is where we observe large discrepancies for the specific heat capacity of different methods. It is in this temperature range that many-body effects can drastically alter the behavior of a system, causing significant deviations between exact and approximate methods. 
An example of this is the LiF system (\reffig{fig:Cv2}(e)).
For this system, the FDHF method predicts, and generally agrees with, the DMQMC  method for the smallest peak at $T\sim 13$. However, with decreasing $T$, the FDHF method predicts a single large peak around $T\sim0.3$ Ha, whereas the DMQMC method predicts a different peak around $T\sim0.6$ Ha with a smaller magnitude and several more peaks appearing as the $T$ decreases.
The peak around $T\sim 13$ in LiF corresponds with the core (1s) electrons from fluorine occupying the higher-energy orbitals such as the 2p orbitals and 2s/2p linear combinations. The second peak for FDHF around $T\sim0.3$ likely corresponds to an overlap of several different electrons occupying higher energy orbitals. In addition to the bonding electrons between the lithium and fluorine atoms, for fluorine, these include the 2s and non-bonding 2p electrons, and for lithium, these include the core 1s electrons.
In FCI, the energy spectrum of LiF spreads out considerably which gives rise to the appearance of more peaks in the DMQMC specific heat capacity.
In general, there are states which have significantly more correlation energy than the ground state, which presumably comes from the ground-state/zero temperature HF orbitals being a less good match to the higher lying energy states. 

For LiF, we saw that FDHF had too few peaks.
For the remainder of the benchmark systems, the FDHF method contains the same number of peaks as the THF and DMQMC methods.
At high temperature, where the lowest energy electrons are primarily excited to higher energy orbitals, the three methods have similar temperatures for the peak in the specific heat capacity.
These low energy electron peaks in the high temperature range of the specific heat capacity are seen in Be (\reffig{fig:Cv2}(a)), BeH$_2$ (\reffig{fig:Cv2}(b)), LiF (\reffig{fig:Cv2}(e)), and N$_2$ (\reffig{fig:Cv2}(f).
The same high temperature peaks are absent from CO (\reffig{fig:Cv2}(c)) and HCN (\reffig{fig:Cv2}(d), where we used the frozen core approximation for the inner shell (1s) electrons for non-hydrogen atoms.\cite{hosteny_ab_1975,werner_molpro_2019}
For those orbitals which are frozen, the orbitals remain occupied regardless of the electronic configuration.
As such, the energy contributions due to the frozen orbitals come from the interaction between the frozen electrons and other electrons able to be excited by temperature.
Therefore, we do not observe direct features resulting from the excitation of these electrons at high temperature, and the frozen orbitals will only indirectly contribute to the specific heat capacities.

The most notable discrepancy between FDHF and the THF and DMQMC methods is the temperature where the specific heat capacity peaks in magnitude at intermediate to low temperatures.
For intermediate to low temperature, which correspond to intermediate to high energy electrons, the FDHF method tends to have the peak in the specific heat capacity at greater temperatures than either THF or DMQMC.
This behavior is generally seen for all systems shown in \reffig{fig:Cv2} excluding LiF which can be considered present or absent depending on the peaks being compared.
Our interpretation of DMQMC and THF having peaks at smaller temperatures is that the many-body effects smooth out the energy gaps between bonding and antibonding orbitals constructed with the 2s and 2p orbitals, leading to a broader peak as each orbital begins to occupy higher energy orbitals in a more gradual fashion, due to the smoothing of energy gaps.
The broadening in the energy of orbitals also results in a broader peak in the specific heat capacity which is why the maximum specific heat capacity is smaller in DMQMC and THF than FDHF.

Though the FDHF has variable performance compared to DMQMC for the systems we tested in \reffig{fig:Cv2}, what is most surprising is the performance of THF in computing specific heat capacities, which gives good estimates for many systems.
In N$_2$ (\reffig{fig:Cv2}(f)), for example, THF is indistinguishable from DMQMC. 
More generally, the location and, to a lesser extent, the magnitude of the peaks for THF are in excellent agreement with the DMQMC method. 
The greatest deviation we observe for this trend is in the LiF system (\reffig{fig:Cv2}(e)), where THF is able to capture one more additional peak in the specific heat capacity over FDHF, but misses the location in $T$ and fails to identify the last peak found at $T\sim0.05$ in DMQMC.

In general, the FDHF method is found to perform qualitatively similar to the DMQMC method for most systems, but falls short in several cases when many-body effects are significant. The ability of THF to close the gap between FDHF and DMQMC specific heat capacities for most systems is due to its ability to include the changes in orbital energies as many-body states are reoccupied. 
The DMQMC method can be seen to be robust in its calculation of heat capacities and quantitative accuracy. 

As a final test for our GPR model of finite temperature data, we calculated the entropies for the approximate methods in \reffig{fig:Cv2} using \refeq{eq:sosentropy}, and compared them with GPR predictions from DMQMC. Figure \ref{fig:entropy2} shows the three methods in comparison with each other. 

In the high $T$ limit, all methods shown in \reffig{fig:entropy2} have nearly indistinguishable entropy, which is what we expect given our use of finite basis sets and fixed particle numbers. Moving down in temperature from high $T$, most of the intermediate $T$ range is observed to be similar between all the methods. It is only in the lower part of the temperature range (generally $T<<1.0$~Ha) when the entropies calculated by the various methods begin to differ.
Overall, we observe the FDHF and THF entropies to be of similar quality to their specific heat capacities that we discussed in \reffig{fig:Cv2}. 
DMQMC generally reproduces the exact entropy; FDHF generally has the most error. The performance of THF lies somewhere in between the other two methods, and is generally of high quality. 
At some lower temperatures, however, the THF entropy shows deviations from the exact entropy. 
Some examples of this are seen in LiF (\reffig{fig:entropy2}(e)), CO (\reffig{fig:entropy2}(c)) and all the hydrogen chain systems (\reffig{fig:entropy2}(g-i)). 
Very occasionally, the THF entropy is more accurate than DMQMC for a small $T$ range, such as in Be (\reffig{fig:entropy2}(a)) and N$_2$ (\reffig{fig:entropy2}(f)). We can attribute this to sampling effects at low $T$ for DMQMC. 

FDHF generally has the most error.
There are several systems where it overestimates the entropy in the intermediate temperature regime, such as in Be (\reffig{fig:entropy2}(a)) around $T\sim2.0$ Ha and LiF (\reffig{fig:entropy2}(e)) around $T\sim0.5$ Ha.
For all systems at low temperature, FDHF underestimates the entropy.
Though FDHF can overestimate the entropy, the underestimating appears more sever as the behavior occurs over a much greater range of temperatures and shows a greater deviation in magnitude from from exact.
A promising observation for DMQMC is that the error in the entropy estimate compared to exact is generally smaller than the error in the FDHF entropy.

Overall, we find the entropy estimates from the approximate methods in \reffig{fig:entropy2} to be of a similar quality to the specific heat capacities discussed in \reffig{fig:Cv2}. Though the approximate methods do much in the way of qualitatively and quantitatively capturing the exact entropy, especially THF, we find that the DMQMC entropy is consistently in the best agreement with the exact specific heat capacity to a degree the approximate methods are unable to match.

\subsection{Beyond ftFCI: methane and water}

We have found that DMQMC in combination with a GPR model allows for the accurate prediction of the specific heat capacity and electronic entropy. 
Having validated this method, we now move on to systems that cannot be treated exactly: water and methane. 
We used a similar protocol as outlined for the DMQMC/GPR algorithm, with several additional modifications. 
The original data set had initiator PIP-DMQMC for $\beta=1$ to $\beta=30$. 
We needed to use the initiator adaptation of DMQMC to converge the energies due to the plateau height of these two systems. 
To this, we added an additional initiator DMQMC calculation between $\beta=0$ and $\beta=1$ with the same input parameters, except that more walkers were added to account for the loss of walkers in DMQMC after the start of a simulation. 
The initiator parameters used were an initiator threshold of $n_{\mathrm{add}}=3.0$ and an initiator level of $n_{\mathrm{ex}}=2$.
For the investigation and discussion of initiator error convergence with respect to walker population we refer readers to Ref.~\onlinecite{vanbenschoten_piecewise_2022}.
To account for the additional information and the step between the DMQMC and PIP-DMQMC data sets, we found it necessary to make two modifications to our methodology.
First, we add the product of an RBF kernel and a Matern 5/2 kernel to our GPR model.
Second, we used the individual simulations for each calculation rather than the average energy, resulting in repeated energies for each $\beta$.

\begin{figure}
\begin{center}
\subfigure[\mbox{}]{\includegraphics[width=0.48\textwidth, height=\textheight, keepaspectratio]{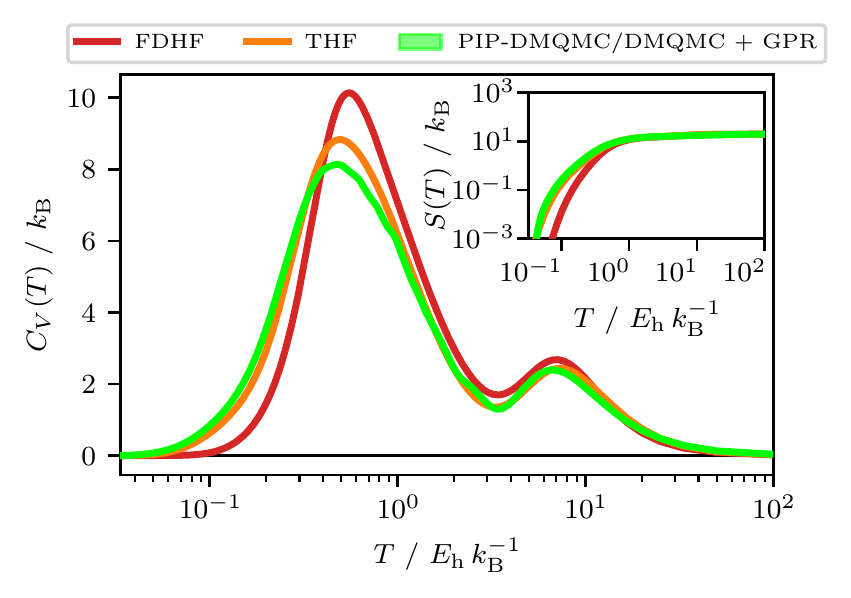}
\label{fig:water}}
\subfigure[\mbox{}]{\includegraphics[width=0.48\textwidth, height=\textheight, keepaspectratio]{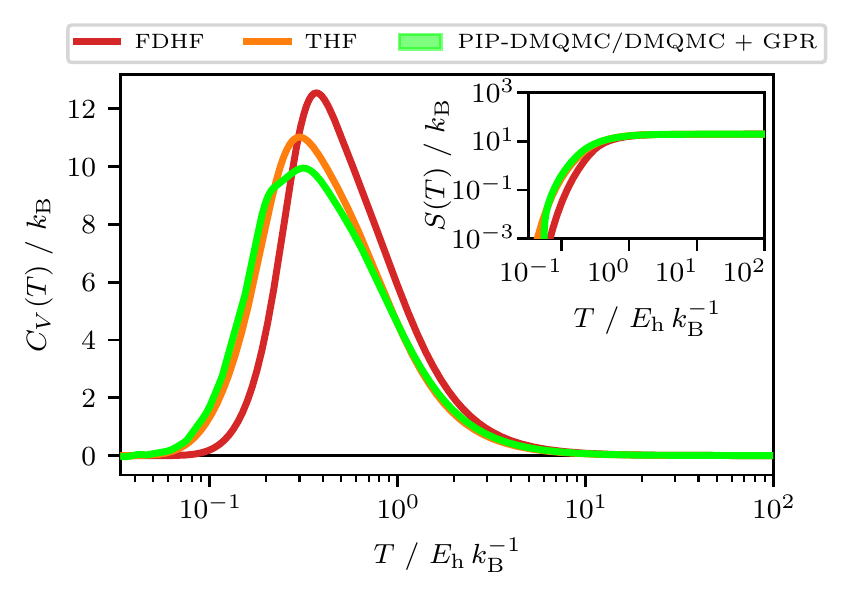}
\label{fig:methane}}
\caption{The specific heat capacity ($C_V$), and entropy ($S$) shown in the inset, calculated as a function of temperature is compared for PIP-DMQMC, THF, and FDHF methods for (a) water and (b) methane.}
\label{fig:pip}
\end{center}
\end{figure}

Figure \ref{fig:pip} shows the specific heat capacities and entropies (inset) from PIP-DMQMC augmented with DMQMC data, for the water molecule in a cc-pVDZ basis set (\reffig{fig:water}) and the methane molecule in a cc-pVDZ basis set (\reffig{fig:methane}). This figure also include the THF and FDHF calculations, so that comparisons can be made to PIP-DMQMC.

For the water molecule (\reffig{fig:water}), we observe that all three methods have similar behavior at either end of the temperature range, approaching zero as we expect. 
All three methods also predict two peaks in the specific heat capacity. The peak at high temperature likely corresponds to excitations coming from the core electrons of the oxygen atom (as these were included for this calculation).
There is a slight but noticeable difference in temperature where the peak occurs for each method.
FDHF has a peak slightly below THF which has a peak near $T\sim 7$, and PIP-DMQMC just below FDHF.
The shift down in temperature where the peak occurs between FDHF and THF is unexpected based on our observations from \reffig{fig:Cv2}, however such behavior may have been previously present to a less noticeable degree.
Most noticible is the increase in magnitude of the FDHF peak, which we had observed previously in BeH$_2$.
As the temperature is decreased, the second peak in the $C_V$ shows a similar effect to the trends observed for the benchmark systems.
Namely that as the accuracy of the method increases, the peak in the specific heat capacity occurs at smaller temperatures with a smaller magnitude.

Finally, we have the entropy for water from all three methods, shown in the inset of \reffig{fig:water}. In this, we can see that all three methods approach the same entropy at high temperature, for the scales used in the figure. The FDHF entropy is found to be much smaller at low temperatures compared to either PIP-DMQMC or THF. Comparing the entropy from PIP-DMQMC to THF, we observe that for most of the temperature range shown, the PIP-DMQMC entropy is higher than or equivalent to the THF entropy. This observation is consistent with all systems in \reffig{fig:entropy2}, except the N$_2$ dimer. With the entropies being nearly identical at high temperatures, and the consistent behavior in both the specific heat capacity and entropy to \reffig{fig:Cv2} and \reffig{fig:entropy2}, we conclude that the PIP-DMQMC data fit with GPR are representative of the exact specific heat capacity and entropy for the water molecule.

For methane (\reffig{fig:methane}), we find that the performance of the PIP-DMQMC and approximate methods is similar to what was observed for water and our benchmark systems. 
That is, in the high and low temperate limits, all three methods agree, with the largest deviations between each method occurring at intermediate temperatures. In addition, all three methods show a single peak. We believe this is reasonable: because the carbon atom employed a frozen core approximation, the remaining electrons are very similar in energy, leading to a single broad peak as these electrons begin occupying higher energy orbitals at similar temperatures.

Comparing the three peaks, the FDHF method has the largest peak in the specific heat capacity around $T\sim0.4$, with a magnitude larger than either PIP-DMQMC or THF, and the FDHF specific heat capacity decays to zero at larger temperatures. The location in temperature for the PIP-DMQMC and THF peaks are very similar.
Comparing the magnitude of the PIP-DMQMC peak to THF, PIP-DMQMC is found to be smaller than THF. This results in the PIP-DMQMC specific heat capacity decaying to zero slower than the THF specific heat capacity. These observations are consistent with the observations we made on water molecule, and our general observations of the benchmark systems in \reffig{fig:Cv2}.

Lastly, we examine the entropy for methane, shown in the inset of \reffig{fig:methane}. The behavior is similar to water: the FDHF entropy is smaller than either PIP-DMQMC or THF at lower temperatures, but all three methods are roughly equivalent at high temperatures for the scale used, which is what we expect. Furthermore, PIP-DMQMC is generally higher than or equivalent to the entropy from THF for the range of temperatures used.

From our analysis of the water and methane molecules simulated in a cc-pVDZ basis set (\reffig{fig:pip}), we find that the PIP-DMQMC specific heat capacities and entropies are generally aligned with our prior observations for the benchmark systems. Therefore, we believe that the GPR method for calculating the specific heat capacity and entropy is able to accurately model the exact specific heat capacity and entropy, even for large systems beyond what can be treated using exact methods. This shows promise for investigating these quantities for large systems beyond exact deterministic treatments.

\begin{figure}
\includegraphics[width=0.45\textwidth,height=\textheight,keepaspectratio]{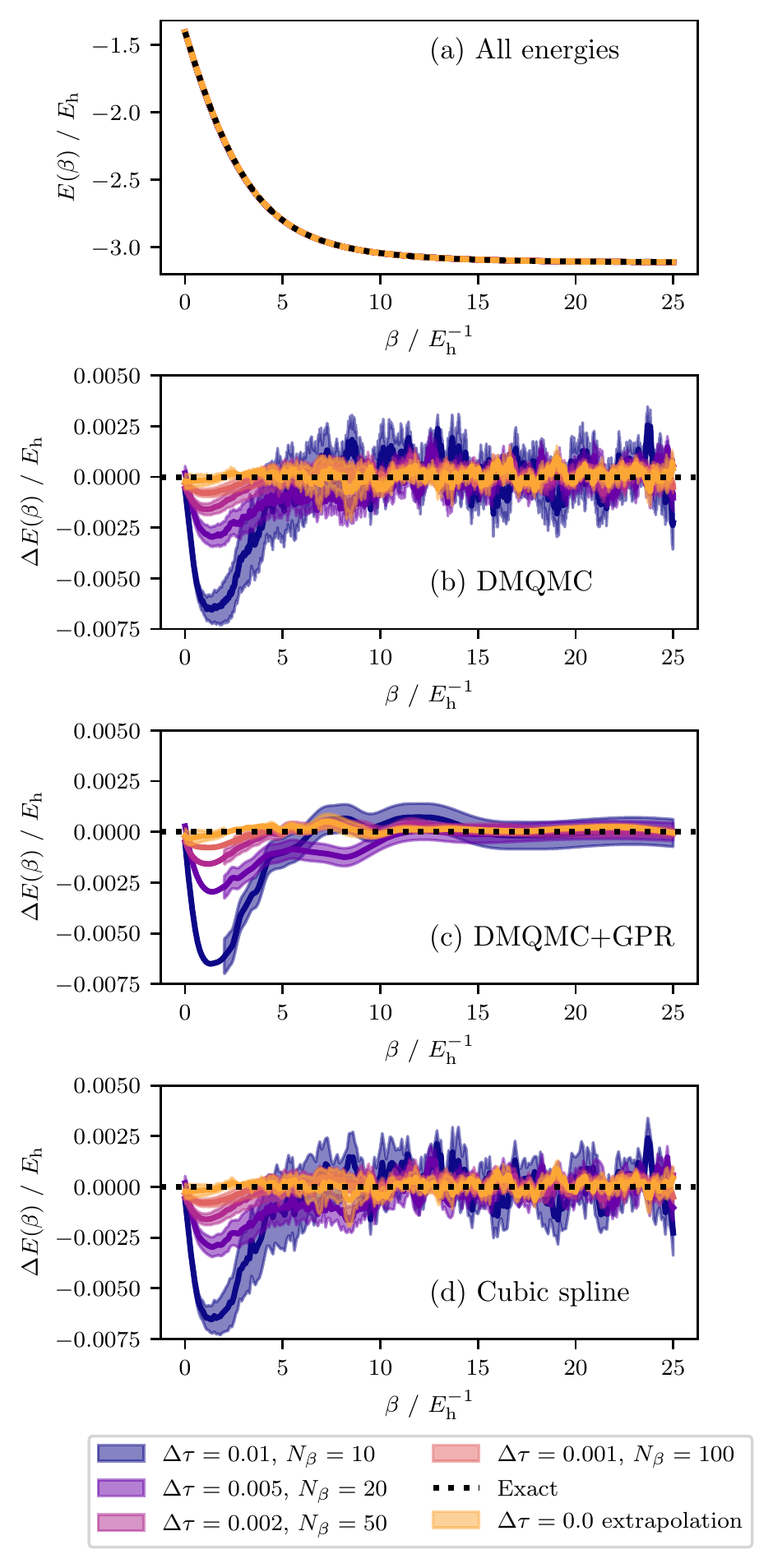}
\caption{A figure showing GPR and cubic spline predictions of finite temperature energies, trained with DMQMC data for the stretched H$_6$/STO-3G system, for a range of $\Delta\tau$ and $N_\beta$. For comparing the quality of the energy estimates on different scales, we show (a) all energies overlaid, (b) the DMQMC energy difference to exact, (c) the GPR predicted energy difference to exact, and (d) the cubic spline predicted energy difference to exact. To generate the GPR predictions we used the same method used to generate GPR models in \reffig{fig:Cv}. The error bars shown in (a) and (d) for the cubic spline energies use the errors from the DMQMC energy for the corresponding $\beta$. The error bars shown in (c) for GPR use the variance returned when generating energy predictions with GPy. For the energies in (c) generated using the $\beta$ fit, the error is smaller than the line and is not visible on the scales used.}
\label{fig:tau}
\end{figure}

\subsection{Changing \texorpdfstring{$\tau$}{time step} and \texorpdfstring{$N_\beta$}{the number of sampling loops}}

One of the effects of the GPR and cubic spline models is that we have been able to smooth over noisy data. 
In general, we noted that we had to downsample the original DMQMC data to improve the fits.
This implies that we might be able to collect fewer data points in the first place. 
We briefly investigated the effect of decreasing the cost of data collection in two different ways: increasing the time step used by up to a factor of 10, and decreasing the number of beta loops we use by up to a factor of 10. 
Each of these independently reduces the calculation cost by a factor of 10. 
We then interpolated over the data set using GPR and cubic splines. 
The results of this are shown in \reffig{fig:tau}. 
These data show that the interpolation matches the original data set. 
The cubic splines generally fit the stochastic noise from the original data, while the GPR fit smooths some of this out. 
The interpolated models \emph{also} reproduce the systematic error in the DMQMC at intermediate $\beta$, referred to as shouldering, which increased with decreasing time step. 

While this is what IP-DMQMC\cite{malone_interaction_2015} (and PIP-DMQMC\cite{vanbenschoten_piecewise_2022}) were designed to overcome, here, our interpolations allow us to attempt something different. 
The shouldering error can be seen as a kind of time-step error\cite{blunt_density-matrix_2014}, which is present in DMQMC but absent in the ground-state method FCIQMC.  The error scales linearly in the time step
(for more details see Appendix \ref{appendix:time_step_error}).
Using the additional interpolated data, we found that it was possible to perform a linear extrapolation on the remaining error. The results are shown in \reffig{fig:tau}, with the result that the timestep error is successfully removed within error for all $\beta$. 
We believe this is promising first step towards manipulating the timestep within DMQMC calculations to make them more efficient.

\section{Conclusions}

In this work, we have introduced a Gaussian Process Regression methodology for fitting finite-temperature DMQMC and PIP-DMQMC energy data to predict specific heat capacities with minimal error. The GPR model requires no prior information about the system---only supervised training.
Comparisons with two common numerical methods used to calculate gradients, finite differences and cubic splines, demonstrate that the DMQMC-predicted gradients are able to accurately capture the behavior of the exact specific heat capacity across the range of data used to train the model.
Thereafter, using numerical integration, we were also able to calculate the electronic entropy, and found the DMQMC entropy produced a consistent behavior across temperatures. DMQMC often captured the exact behavior for many of the temperatures investigated---a result that the finite difference and cubic spline methods were not always able to achieve.
We found that the DMQMC specific heat capacities and entropies were qualitatively and quantitatively distinct from those predicted by the THF and FDHF methods, which are deterministic and approximate methods.
As a final test of the GPR fits of finite temperature DMQMC and PIP-DMQMC data, we applied the GPR method to two large systems that we do not have exact data for, the water and methane molecules. We found that GPR performed well on these systems.
We believe that our results suggest that the GPR methodology outlined here will work with other finite temperature QMC which sample the finite temperature energy reasonably well.\cite{zhang_finite_1999,shen_finite_2020,he_reaching_2019,he_finite-temperature_2019,liu_ab_2018,yilmaz_restricted_2020,dornheim_ab_2021,brown_path-integral_2013,dornheim_permutation_2015,militzer_development_2015,larkin_phase_2017,groth_configuration_2017,dornheim_ab_2018,dornheim_fermion_2019,yilmaz_restricted_2020,chang_recent_2015}
As such, we hope others find the data presented here useful for comparison and benchmarking the application of GPR to find gradients of noisy data similar to ours.

This manuscript has focused on the benefits to physical modeling of using GPR to interpolate between energy/temperature data points, namely the ability to calculate specific heat capacities and entropies. When it comes to treating noise, our preliminary investigations show improvements in the quality of the cubic spline and GPR fits when adding different noise models. For GPR, this includes less dependence on the beta/temperature cross-over point. This is something we plan to study systematically and have left for further work.

Another key limitation of this study is that we only work with molecules and have made a number of approximations: we only consider one geometry (neglecting vibrational and rotational effects), one symmetry, no continuum states, and finite basis sets.
As such, the entropies and electronic heat capacities that we calculate are only intended to be a proof of concept, rather than for systems with a practical relevance.
These applications do exist, for example conductive solids, and moving in that direction is a long-term goal of ours.
The success of GPR shown in this paper bodes well for its eventual use in these applications.
A final limitation of our study is that we did not investigate a method for systematically determining the best kernel combinations, which GPR can be sensitive to\cite{rasmussen_gaussian_2006}, or crossover between the $T$ and $\beta$ fits to use during the regression (and under what circumstances our choices are insufficient). 
Still, we believe that our GPR methodology for fitting DMQMC data will be able to provide access to specific heat capacities and entropies for even larger systems than the methane and water molecules studied in this work, and that this proof of concept will be of interest to the broader community.

\section{Acknowledgements}

Research was supported by the U.S. Department of Energy, Office of Science,
Office of Basic Energy Sciences Early Career Research Program (ECRP) under
Award Number DE-SC0021317. 

This research also used resources from the University of Iowa and the resources
of the National Energy Research Scientific Computing Center, a DOE Office of
Science User Facility supported by the Office of Science of the U.S. Department
of Energy under Contract No. DE-AC02-05CH11231 (computer time for calculations
only) using NERSC award BES-ERCAP0019952.

For the purposes of providing information about the calculations used, files
will be deposited with Iowa Research Online (IRO) with a reference number [to
be inserted at production].

We gratefully acknowledge Michael Mavros for written comments on the manuscript.

\section{Author contributions}

\contributions

\section{Data Availability}

The data that supports the findings of this study are available within the
article. 

\section{Conflicts of Interest}

The authors have no conflicts to disclose.

\appendix

\section{The DMQMC algorithm}
\label{appendix:dmqmc_algo}

The spawning, cloning, and annihilation steps were originally developed for the ground state FCIQMC method, of which DMQMC is a finite-temperature generalization.
In this work, we used real walker amplitudes with a spawning cutoff of $p_{\mathrm{cut}}=0.01$, which is reflected in the spawning step by stochastically rounding spawns below this value to the cutoff or zero.\cite{petruzielo_semistochastic_2012}
The process begins by looping through the occupied sites $f_{ij}(\tau)$, with a walker population $N_{w,ij}(\tau)=|f_{ij}(\tau)|$.
The spawning and cloning steps are carried out for each site. Annihilation occurs once spawning and cloning has completed for all sites. The details of each step are:
\begin{enumerate}
\item[] \textit{Spawning:} For each walker on $f_{ij}(\tau)$, spawning is attempted. A total of $n_s=\mathrm{floor}(N_{w,ij}(\tau))$ attempts are made, where we define $\mathrm{floor}(x)$ as the integer component of $x$. Additionally, another attempt is made if the condition $N_{w,ij}(\tau)-n_s>r$ is met, where $r$ is a random number uniformly selected in the range from inclusive zero to noninclusive one ($r\in\left[0,1\right)$). During each attempt, another site $f_{kj}$ is randomly selected such that $k\ne i$, by generating an allowed random excitation between $|D_i\rangle$ and $|D_k\rangle$. The probability of the excitation is given as $p_{\mathrm{gen}}$. Then given $p_s=\frac{\Delta\tau}{2 p_{\mathrm{gen}}}|H_{ik}|$, if the condition $p_s>p_{\mathrm{cut}}$ is met, $p_s$ walkers are spawned. Otherwise if the condition $p_s > r\times p_{\mathrm{cut}}$ is met, $p_{\mathrm{cut}}$ walkers are spawned. If no condition is met, no walkers are spawned. The sign of spawned walkers if given by: $q_{kj}=\mathrm{sign}(-H_{ik}f_{ij})$, where we define $\mathrm{sign}(x)=1.0$ if $x>0.0$ and $-1.0$ if $x<0.0$. In the same attempt, a similar process is carried out for a randomly selected site $f_{ik}$. For successful spawns, decimals outside the tolerance of the method are stochastically rounded to the tolerance.
\item[] \textit{Cloning:} Walkers are either cloned or removed from $f_{ij}(\tau)$. The number of walkers cloned or removed is: $p_{c}=f_{ij}(\tau)\frac{\Delta\tau}{2}(2S-H_{ii}-H_{jj})$. The new walkers are cloned if $\mathrm{sign}(f_{ij}(\tau))=\mathrm{sign}(p_{c})$, and removed when $\mathrm{sign}(f_{ij}(\tau))\ne\mathrm{sign}(p_{c})$. The cloning/removal occurs immediately.
\item[] \textit{Annihilation:} Walkers created in the \textit{spawning} step are added to their appropriate site, and walkers with opposite signs on the same site are annihilated leaving only a single sign of walkers. For any site with a walker magnitude below one, the walkers are stochastically rounded to one, with the same sign, or zero.
\end{enumerate}
Here $p_{\mathrm{gen}}$ is calculated following the original FCIQMC procedure described by Booth \textit{et. al.},\cite{booth_fermion_2009} though we do note there are more efficient algorithms for generating excitations which have not yet been tested in DMQMC.\cite{neufeld_exciting_2018}

These steps are computationally efficient, as only a few spawning attempts are required per walker; additionally, the memory cost is minimized by only storing the largest elements of the density matrix.
We note that the symmetric Bloch equation (\refeq{eq:bloch2}) was used above, but the asymmetric Bloch equation can also be used
\begin{equation}
\frac{\mathrm{d}}{\mathrm{d}\tau}\hat{f}(\tau)=-\hat{f}(\tau)\hat{H}.
\label{eq:asym_bloch}
\end{equation}
The only changes required to accommodate the asymmetric form are: \textit{spawning} now occurs only between $f_{ij}$ and $f_{ik}$, the factor $\frac{\Delta\tau}{2}$ becomes $\Delta\tau$, and the \textit{cloning} term is now $p_c=f_{ij}(\tau)\Delta\tau(S-H_{jj})$.

\section{The initiator approximation}
\label{appendix:initiator}

To overcome issues resulting from the sign problem, a system dependent minimum number of walkers, referred to as the `plateau', is required when using FCIQMC or DMQMC.
When DMQMC was originally introduced, it was posited the DMQMC plateau will scale as the square of the FCIQMC plateau.
However, in our investigation of the DMQMC sign problem, we found that several factors such as propagator symmetry and the initial condition can influence the DMQMC plateau.
Based on these factors, the DMQMC plateau can scale anywhere from linear to the square of the FCIQMC plateau.\cite{petras_sign_2021}
For small systems, the computational cost of overcoming the plateau is easily achieved on modern computers.
However, for large systems, the number of walkers can exceed the computational resources available on most modern computers.

To overcome the sign problem, the initiator approximation is used, which was originally developed for FCIQMC.\cite{cleland_communications_2010}
Thereafter, the initiator approximation was adapted for DMQMC methods by Malone and coworkers.\cite{malone_accurate_2016}
The initiator approximation places restrictions on spawning to unoccupied sites.
If no initiator criteria are met, the spawns produced in the \textit{spawning} step are set to zero.
These criteria are:
\begin{itemize}
\item The population ($N_{w,ij}(\tau)$) on $f_{ij}$ must be greater than or equal to a user defined threshold $n_{\mathrm{add}}$ ($N_{w,ij}(\tau) \ge n_{\mathrm{add}}$). In this work we used $n_{\mathrm{add}}=3.0$.
\item The number of excitations between the site labels $|D_i\rangle$ and $|D_j\rangle$ for $f_{ij}$ is less than or equal to a user defined excitation threshold $n_{\mathrm{ex}}$. In this work we used $n_{\mathrm{ex}}=2$.
\item If two (or more) spawns arriving to a site from separate sites have the same sign, the spawn is allowed.
\end{itemize}
The criteria imposed during the \textit{spawning} step introduce a systematically improvable error in the energy, which is removed as $N_w(\tau)$ increases.\cite{cleland_communications_2010,malone_accurate_2016,cleland_study_2011,cleland_taming_2012,shepherd_investigation_2012,booth_towards_2013,thomas_accurate_2015} The initiator error can be removed by increasing $N_w(\tau)$ until the energy remains unchanged.
Additionally, there has been continued work within FCIQMC to remove the error introduced by the initiator approximation.\cite{ghanem_adaptive_2020,ghanem_unbiasing_2019}
Such developments have not been tested in DMQMC, but could reasonably provide similar improvements to those seen in FCIQMC.

\section{Finite difference}
\label{appendix:finite_difference}

The finite difference for calculating the approximate derivative of evenly spaced data is
\begin{align}
\frac{\Delta f(x_i)}{\Delta x}=
\begin{cases}
\frac{1}{\Delta x}(f(x_{i+1})-f(x_{i})), & i = 1, \\
\frac{1}{2\Delta x}(f(x_{i+1})-f(x_{i-1})), & 1 < i < N, \\
\frac{1}{\Delta x}(f(x_{i})-f(x_{i-1})), & i = N,
\end{cases}
\label{eq:finite_difference}
\end{align}
where $f(x)$ is the dependent variable corresponding to the independent variable $x_i\in\{x_1,x_2,...,x_N\}$, and $\Delta x$ is the gap between evenly spaced independent variables.

Computing finite differences is a resource-efficient method that does not rely on the availability of an analytical functional form for an expression, only requiring well-sampled numerical data. 
This is advantageous when direct access to an analytical functional expression is not practical or possible, such as in DMQMC. 
Finite differences have the potential to work well because DMQMC data sets are large and data points can be collected with a consistent spacing in $\beta$. 
The main downside to using finite difference methods with DMQMC data is the stochastic error present in the energy estimates.\cite{petras_using_2020}
As $\beta$ becomes larger, the change in the energy with respect to each timestep $\Delta\tau$
can become smaller than the stochastic error, and can lead to undesirable outcomes in the finite difference. 
Additionally, using a finite difference method for our purposes/specific purpose does not generate a smooth or continuous energy function.

\section{Cubic spline}
\label{appendix:cubic_spline}

Cubic splines are useful because a numerical dataset can be used to generate an analytical function that is continuously differentiable on the domain of the original dataset. 
One advantage to having a continuous function for the energy is that there is no restriction on the $x$ values used within the domain of the original dataset.
In the context of this current paper, cubic splines are the current way that energy data can be interpolated for the purposes of finding specific heat capacities\cite{liu_unveiling_2020}.
Cubic splines are computationally efficient and numerically robust, which allows for rapid iteration through different parameters. 
A disadvantage of cubic splines when compared to other methods is that it is not possible to extrapolate beyond the domain of the original data using splines alone. 
Cubic spline interpolation is also quite sensitive to noise, which is important because the DMQMC data has stochastic error that can cause problems in correctly identify the underlying analytical function. 
Cubic splines can also introduce artifacts due limitations in the form of the function. 

\begin{figure}
\includegraphics[width=0.48\textwidth, height=\textheight, keepaspectratio]{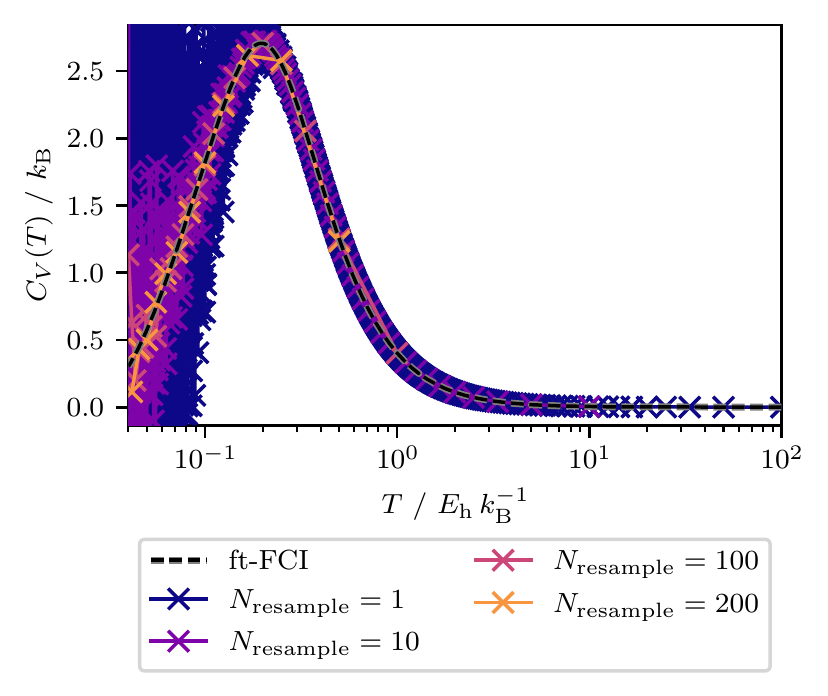}
\caption{Calculations on the stretched H$_6$/STO-3G system showing cubic spline calculations of the specific heat capacity ($C_V$) with different sampling frequencies for the DMQMC data set.
Geometry details for the H$_6$ system, and simulation details for DMQMC, are located in the calculation details (\refsec{sec:calc}).}
\label{fig:opt}
\end{figure}
\begin{figure*}
\includegraphics[width=\textwidth,height=\textheight,keepaspectratio]{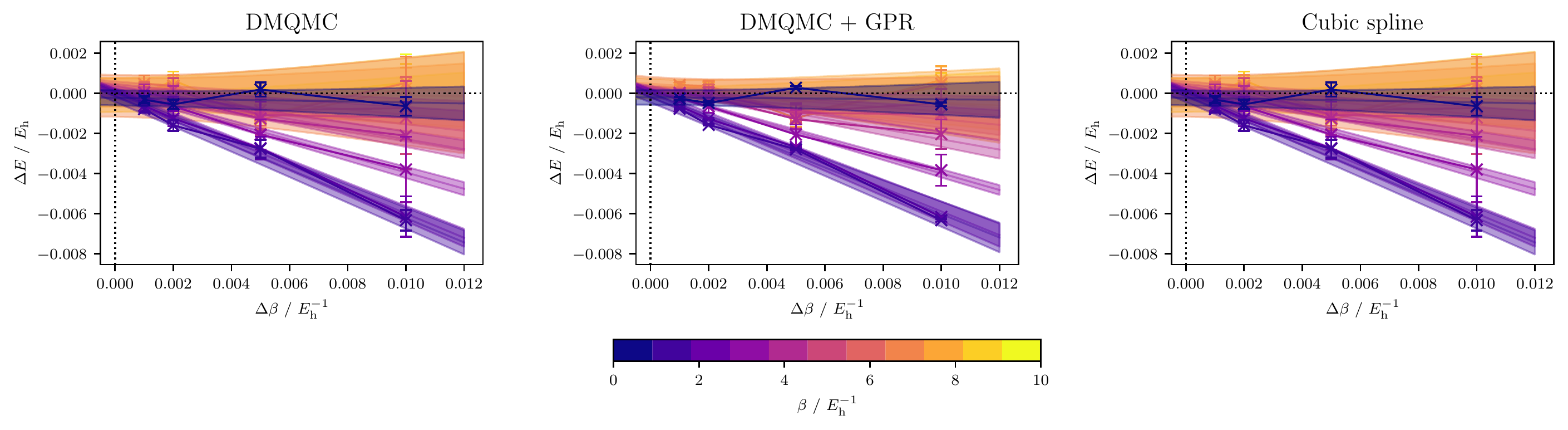}
\caption{The time step error ($\Delta E$) 
for H$_6$/STO-3G is plotted for different time steps. The value $\Delta E$ is calculated by taking the difference between the energy labelled at the top of the plot and ft-FCI. A time step error of zero ($\Delta E=0$) can be achieved by a linear extrapolation.}
\label{fig:timestep}
\end{figure*}

Figure \ref{fig:opt} shows four attempts at calculating the specific heat capacity for stretched H$_6$ using different re-sampling intervals.
These each have different appearances in the low-$T$ regime. 
For $C_V$, the growth of noise in the low-$T$ regime comes because the points are equally spaced in $\beta$ (thus causing the spacing in $T$ to decrease) at the same time as the energy decreases as $\beta$ increases. 
The noise is, therefore, worse when the re-sampling interval is smaller because the change in $\beta$ is smaller; as the re-sampling interval increases, the results get better.
However, this has the effect of eventually missing some of the features of the $C_V$ peak, which can be seen in the data where one in every 200 data point is taken.

When we looked at this over more systems, we found that, at the smallest re-sampling interval, the failure due to the noise in the $C_V$ curve is severe. However, at the largest re-sampling interval, the important features of the $C_V$ are not always included. 
This balance between too much and too little resampling was also found in our entropy calculations (for which details appear above in this manuscript). 
Therefore, we find that a re-sampling of every one out of ten points is most consistent with the GPR methodology, and achieves good performance, so it is used for all systems.

\section{Time step error}
\label{appendix:time_step_error}

Figure \ref{fig:timestep} shows the linear extrapolations used to produce \reffig{fig:tau}. %
Each data series/color represents a separate $\beta$ value and the linear extrapolation into the origin demonstrates the trend that the time step error is linear in the time step. 
Extrapolations in the middle of the range of $\beta$ show a linear trend the most clearly. 
These data show that the time step error is linear in $\Delta\tau$. 
The magnitude of the uncorrected time step error can be seen to be lowest for high and for low $\beta$, and the extrapolation line for these $\beta$ values is flat.

% \bibliography{wzv}


 \end{document}